\pdfoutput=1
\documentclass[11pt,twoside,a4paper,cmspaper,final,collab]{cms-tdr}
\def\svnVersion{ea80a74}\def\svnDate{2023/10/12}\def\cmsCernNoTag{CERN-EP-2023-061}\def\cmsCernDate{\today}\def\cmsMessage{Published in Physical Review D as \href{http://dx.doi.org/10.1103/PhysRevD.108.072004}{\doi{10.1103/PhysRevD.108.072004}.}}
\begin{document}\cmsNoteHeader{HIG-22-002}

\newcommand{\mb}{\mbox{\ensuremath{{\,\text{mb}}}}\xspace}
\newcommand{\ggH}{\ensuremath{{\Pg\Pg\PH}}\xspace}
\newcommand{\Yemu}{\ensuremath{{\abs{Y_{\Pe\PGm}}}}\xspace}
\newcommand{\Ymue}{\ensuremath{{\abs{Y_{\PGm\Pe}}}}\xspace}
\newcommand{\Hem}{\ensuremath{{\PH\to\Pe\PGm}}\xspace}
\newcommand{\HemSM}{\ensuremath{{\PH\to\Pe\PGm}}\xspace}
\newcommand{\Xem}{\ensuremath{{\PX\to\Pe\PGm}}\xspace}
\newcommand{\Het}{\ensuremath{{\PH\to\Pe\PGt}}\xspace}
\newcommand{\Hmt}{\ensuremath{{\PH\to\PGm\PGt}}\xspace}
\newcommand{\BHemSM}{\ensuremath{{\mathcal{B}(\PH\to\Pe\PGm)}}\xspace}
\newcommand{\ppXem}{\ensuremath{{\Pp\Pp \to \PX \to \Pe \PGm}}\xspace}
\newcommand{\ppXemOFS}{\ensuremath{{\Pp\Pp \to \PX(146) \to \Pe \PGm}}\xspace}
\newcommand{\sppXem}{\ensuremath{{\sigma(\Pp\Pp \to \PX \to \Pe \PGm)}}\xspace}
\newcommand{\sppXemOFS}{\ensuremath{{\sigma(\Pp\Pp \to \PX(146) \to \Pe \PGm)}}\xspace}
\newcommand{\mutau}{\ensuremath{{\PGm\PGt}}\xspace}
\newcommand{\emu}{\ensuremath{{\Pe\PGm}}\xspace}
\newcommand{\etau}{\ensuremath{{\Pe\PGt}}\xspace}
\newcommand{\egm}{\ensuremath{{\Pe\PGm}}\xspace}
\newcommand{\Zee}{\ensuremath{{\PZ\to\Pe\Pe}}\xspace}
\newcommand{\Zmm}{\ensuremath{{\PZ\to\PGm\PGm}}\xspace}
\newcommand{\Ztt}{\ensuremath{{\PZ\to\PGt\PGt}}\xspace}
\newcommand{\mem}{\ensuremath{{m_{\Pe\PGm}}}\xspace}
\newcommand{\mh}{\ensuremath{{m_{\PH}}}\xspace}
\newcommand{\mx}{\ensuremath{{m_{\PX}}}\xspace}
\newcommand{\mjj}{\ensuremath{{m_{j_{1}j_{2}}}\xspace}}
\newcommand{\wjets}{\ensuremath{{\text{QCD }\PW\text{+jets}}}\xspace}
\newcommand{\zjets}{\ensuremath{{\PZ\text{+jets}}}\xspace}
\newcommand{\detajj}{\ensuremath{{\abs{\Delta\eta(j_{1},j_{2})}}}\xspace}
\newcommand{\irel}{\ensuremath{{I^\ell_{\text{rel}}}}\xspace}
\newcommand{\irele}{\ensuremath{{I^\Pe_{\text{rel}}}}\xspace}
\newcommand{\irelm}{\ensuremath{{I^\PGm_{\text{rel}}}}\xspace}
\newcommand{\aeta}{\ensuremath{{\abs{\eta}}}\xspace}
\newcommand{\dr}{\ensuremath{{\Delta R}}\xspace}
\ifthenelse{\boolean{cms@external}}{\providecommand{\cmsTable}[1]{#1}}{\providecommand{\cmsTable}[1]{\resizebox{\textwidth}{!}{#1}}}
\ifthenelse{\boolean{cms@external}}{\providecommand{\cmsLeft}{upper\xspace}\providecommand{\cmsLLeft}{Upper\xspace}}{\providecommand{\cmsLeft}{left\xspace}\providecommand{\cmsLLeft}{Left\xspace}}
\ifthenelse{\boolean{cms@external}}{\providecommand{\cmsRight}{lower\xspace}\providecommand{\cmsRRight}{Lower\xspace}}{\providecommand{\cmsRight}{right\xspace}\providecommand{\cmsRRight}{Right\xspace}}
\cmsNoteHeader{HIG-22-002}

\title{Search for the lepton-flavor violating decay of the Higgs boson and additional Higgs bosons in the \texorpdfstring{${\emu}$}{emu} final state in proton-proton collisions at \texorpdfstring{${\sqrt{s} = 13\TeV}$}{sqrt(s) = 13 TeV}}

\author*[cern]{The CMS Collaboration}

\date{\today}

\abstract{
A search for the lepton-flavor violating decay of the Higgs boson and potential additional Higgs bosons with a mass in the range {110--160\GeV} to an $\Pe^{\pm}\PGm^{\mp}$ pair is presented. The search is performed with a proton-proton collision dataset at a center-of-mass energy of 13\TeV collected by the CMS experiment at the LHC, corresponding to an integrated luminosity of 138\fbinv. No excess is observed for the Higgs boson. The observed (expected) upper limit on the $\Pe^{\pm}\PGm^{\mp}$ branching fraction for it is determined to be $4.4\ (4.7)\times10^{-5}$ at 95\% confidence level, the most stringent limit set thus far from direct searches. The largest excess of events over the expected background in the full mass range of the search is observed at an $\Pe^{\pm}\PGm^{\mp}$ invariant mass of approximately {146\GeV} with a local (global) significance of 3.8 (2.8) standard deviations.
}

\hypersetup{%
pdfauthor={CMS Collaboration},%
pdftitle={Search for the lepton-flavor violating decay of the Higgs boson and additional Higgs bosons in the emu final state in proton-proton collisions at sqrt(s) = 13 TeV},%
pdfsubject={CMS},%
pdfkeywords={CMS, lepton-flavor violation, higgs}}

\maketitle

\section{Introduction}
The Higgs boson (${\PH}$) was discovered by the ATLAS and CMS experiments at the LHC in 2012 with mass ${\mh\approx125\GeV}$~\cite{Aad:2012tfa,Chatrchyan:2012ufa,Chatrchyan:2013lba}. Measurements of the properties of the Higgs boson, including the \PH decay branching fractions, are thus far found to be consistent with the expectations of the standard model (SM)~\cite{ATLAS:2022Higgs, CMS:2022dwd, Aad:2015gba, Khachatryan:2014jba, Chatrchyan:2012jja, Aad:2013xqa, Khachatryan:2014kca, Sirunyan:2017exp}. Previous studies based on the combined results from the ATLAS and CMS experiments constrain the inclusive branching fraction of potential beyond-the-SM (BSM) undetected visible decays of the Higgs boson to be $<$0.12 and $<$0.16 at the 95\% confidence level (\CL), respectively~\cite{ATLAS:2022Higgs, CMS:2022dwd}. 

The lepton-flavor violating (LFV) decays \Hem, \Het, or \Hmt are forbidden in the SM but may arise in BSM theories with more than one Higgs boson doublet~\cite{Bjorken:1977vt, BRANCO20121}, models with flavor symmetries~\cite{Ishimori:2010au}, the Randall-Sundrum model~\cite{Perez:2008ee, Casagrande:2008hr, Buras:2009ka, Blanke:2008zb, Albrecht:2009xr}, composite Higgs models~\cite{Agashe:2009di, Azatov:2009na}, certain supersymmetric models~\cite{DiazCruz:1999xe, Han:2000jz, Arhrib:2012ax}, and others~\cite{AguilarSaavedra:2009mx, Goudelis:2011un, McKeen:2012av, Pilaftsis:1992st, Korner:1992zk}. In these models, the LFV decays can occur through the off-diagonal LFV Yukawa couplings ${Y_{\egm}}$, ${Y_{\etau}}$, or ${Y_{\mutau}}$, which couple the Higgs boson with leptons of different flavor. The presence of these off-diagonal LFV Yukawa couplings may enhance processes such as ${\PGm\to3\Pe}$, ${\PGm\to\Pe}$ conversion, and ${\PGm\to\Pe\Pgg}$ that could proceed via a virtual Higgs boson exchange~\cite{Shanker:1981mj, McWilliams:1980kj}. In particular, the most stringent limit on \BHemSM is obtained indirectly from the limit on ${\PGm\to\Pe\Pgg}$~\cite{MEG:2016leq} to be ${<}10^{-8}$~\cite{Harnik:2012pb}. However, the indirect limit on \HemSM assumes the SM values for the not yet tightly constrained Yukawa couplings ${Y_{\PGm\PGm}}$~\cite{ATLAS:2020fzp,CMS:2020xwi} and the unmeasured ${Y_{\Pe\Pe}}$. For example, should ${Y_{\PGm\PGm}}$ be smaller than the SM prediction, the indirect constraints on \BHemSM mentioned would be loosened. It also assumes the flavor changing neutral current is dominated by the Higgs boson contribution. Hence, a direct search for \HemSM remains important. The most stringent direct limit on \BHemSM, thus far, was set by the ATLAS experiment at an observed (expected) limit of ${6.2\ (5.9)\times10^{-5}}$ at 95\% \CL with a proton-proton (\Pp{}\Pp) collision dataset at a center-of-mass energy of {13\TeV}, corresponding to an integrated luminosity of {139\fbinv}~\cite{Aad:2690381}. 

LFV could also arise in decays of additional Higgs bosons in the Type-III two Higgs doublet model (2HDM)~\cite{BRANCO20121}. Recent studies have shown that searching for additional Higgs bosons with a mass below twice the \PW boson mass in the LFV decay channels is particularly important to constrain the Type-III 2HDM parameter space~\cite{20202HDM}. An additional Higgs boson with a mass larger than twice the \PW boson mass is expected to decay primarily into a $\PWp\PWm$ pair which dilutes the rate of LFV decays. 

This paper reports a search for a LFV decay in the \emu channel of both \PH and of an additional Higgs boson (\PX) with a mass, \mx, in the range {110--160\GeV}. The upper range of {160\GeV} corresponds to twice the \PW boson mass. The search is performed with data recorded by the CMS experiment in \Pp{}\Pp collisions at a center-of-mass energy of 13\TeV during the period from {2016--2018} and corresponding to an integrated luminosity of {138\fbinv}. The analysis in this paper is optimized for the two dominant production modes of the Higgs boson at the LHC: gluon fusion (\ggH) and vector boson fusion (VBF). The final state of interest in both production modes consists of a prompt, oppositely-charged electron-muon pair. Subdominant production modes of the Higgs boson in association with a vector boson (\PW or \PZ) are not considered due to the smaller cross sections.

This paper is organized as follows: a description of the CMS detector is given in Sec.~\ref{sec:detector}, the collision data and simulated samples are discussed in Sec.~\ref{sec:dataset}, the event reconstruction is described in Sec.~\ref{sec:event_reco}, and the event selection is described in Sec.~\ref{sec:selection}. The event categorization is described in Sec.~\ref{sec:categorization}. Signal and background modeling, and systematic uncertainties are described in Sec.s~\ref{sec:modeling} and~\ref{sec:sysUnc}, respectively. Results are presented in Sec.~\ref{sec:results}, and the summary is given in Sec.~\ref{sec:summary}. 

\section{The CMS detector}
\label{sec:detector}

The central feature of the CMS apparatus is a superconducting solenoid of {6\unit{m}} internal diameter, providing a magnetic field of {3.8\unit{T}}. Within the solenoid volume are a silicon pixel and strip tracker, a lead tungstate crystal electromagnetic calorimeter (ECAL), and a brass and scintillator hadron calorimeter (HCAL), each composed of a barrel and two endcap sections. Forward calorimeters extend the pseudorapidity, ${\eta}$, coverage provided by the barrel and endcap detectors. Muons are detected in gas-ionization chambers embedded in the steel flux-return yoke outside the solenoid. A more detailed description of the CMS detector, together with a definition of the coordinate system used and the relevant kinematic variables, can be found in Ref.~\cite{Chatrchyan:2008zzk}.

Events of interest are selected using a two-tiered trigger system. The first level, composed of custom hardware processors, uses information from the calorimeters and muon detectors to select events at a rate of approximately {100\unit{kHz}} within a fixed latency of approximately {4\mus}~\cite{Sirunyan:2020zal}. The second level, the high-level trigger, consists of a farm of processors running a version of the full event reconstruction software optimized for fast processing that reduces the event rate to approximately {1\unit{kHz}} before data storage~\cite{Khachatryan:2016bia}.

\section{Collision data and simulated events}
\label{sec:dataset}
This search is carried out using \Pp{}\Pp collision data collected by the CMS experiment from {2016--2018} at a center-of-mass energy of 13\TeV with the integrated luminosity being 36.3 in 2016, 41.5 in 2017, and 59.8\fbinv in 2018, respectively. Single-electron or -muon triggers with isolation criteria are used to collect the data. The transverse momentum, \pt, thresholds for the electron (muon) trigger are 27 (24), 32 (27), and {32 (24)\GeV} in the 2016, 2017 and 2018 datasets, respectively.

Simulations are used to model the signal and background events. To model the parton showering, hadronization, and underlying event properties, \PYTHIA~\cite{Sjostrand:2014zea} version 8.240, with the CP5 underlying event tune~\cite{Sirunyan:2019dfx} is used in all cases. The NNPDF3.1 parton distribution functions (PDFs) are used in the simulations~\cite{Ball:2017nwa}. The simulation of interactions in the CMS detector is based on \GEANTfour~\cite{Agostinelli:2002hh}; the same reconstruction algorithms are used as for data.

The Higgs bosons are produced at the LHC predominantly via the \ggH mode~\cite{Georgi:1977gs}, the VBF mode~\cite{Cahn:1986zv}, and in association with a vector boson (\PW or \PZ)~\cite{Glashow:1978ab}. Signal samples of \HemSM and \Xem with a hypothesized \mx of 110, 120, 130, 140, 150, and {160\GeV} are generated for the \ggH and VBF modes at next-to-leading order (NLO) accuracy in perturbative quantum chromodynamics (QCD) with the \POWHEG v2.0 generator~\cite{Nason:2004rx, Frixione:2007vw, Alioli:2010xd, Alioli:2010xa, Alioli:2008tz, Bagnaschi:2011tu} using the implementation described in Refs.~\cite{Heinrich:2017kxx, Buchalla:2018yce}, interfaced with \PYTHIA. The simulated {\PX}s are assumed to have narrow width. The \HERWIG 7.2 generator~\cite{Bellm:2015jjp} with the CH3 underlying event tune~\cite{CMS:2020dqt} interfaced with the \POWHEG v2.0 generator, is used to produce alternative samples for the VBF signal. These samples are used to evaluate the systematic uncertainty in the kinematic distributions of the final state particle in VBF production due to different choices of parton shower simulation~\cite{Jager:2020hkz}.

Background events from \PH decaying to a pair of \PGt leptons are simulated for all three dominant production modes at the LHC at NLO with the same \POWHEG v2.0 generator as the signals, interfaced with \PYTHIA. Background events from \PH decaying to a pair of \PW bosons are generated similarly for the \ggH and VBF modes only as the contribution of other production modes is negligible. 

The \MGvATNLO generator~\cite{Alwall:2014hca} (version 2.6.5) is used to simulate the single \PW/\PZ backgrounds produced by VBF in association with two or more jets from electroweak vertices (VBF \PW/\PZ+jets) at leading order with the MLM jet matching and merging schemes~\cite{Alwall:2007fs}. Drell-Yan (DY), single \PW with jets from QCD vertices (\wjets), and diboson (${\PW\PW}$, ${\PW\PZ}$, ${\PZ\PZ}$) events are simulated with the same generator at NLO, with the FxFx jet-matching and merging scheme~\cite{Frederix:2012ps}. Top quark-antiquark pair and single top quark production are generated at NLO with \POWHEG v2.0.

All samples include the effects of additional ${\Pp\Pp}$ interactions in the same or adjacent bunch crossings, referred to as pileup. The distribution of the number of pileup interactions in simulation is also weighted to match the one observed in data.

\section{Event reconstruction}
\label{sec:event_reco}

The particle flow (PF) algorithm~\cite{CMS-PRF-14-001} reconstructs and identifies particles in an event through an optimized combination of information from the various subdetectors of the CMS detector. The identification of the particle type (photons, electrons, muons, charged and neutral hadrons) plays an important role in determining the direction and energy of each reconstructed particle (PF candidates). The primary vertex (PV) is taken to be the vertex corresponding to the hardest scattering in the event, evaluated using tracking information alone, as described in Sec. 9.4.1 of Ref.~\cite{CMS-TDR-15-02}. 

An electron is identified as a track from the PV combined with one or more ECAL energy clusters. These clusters correspond to the electron and possible bremsstrahlung photons emitted when passing through the tracker. Electrons are accepted in the range of ${\aeta<2.5}$, except for ${1.44<\aeta<1.57}$, the transition region between the barrel and endcap calorimeters, because the reconstruction of an electron object in this region is not optimal. Electrons with ${\pt>10\GeV}$ are identified with an efficiency of 80\% using a multivariate discriminant that combines observables sensitive to the amount of bremsstrahlung energy deposited along the electron trajectory, the geometric and momentum matching between the electron trajectory and the associated clusters, and the distribution of the shower energy in the calorimeters~\cite{CMS:2020uim}. Electrons identified as originating from photon conversions are removed. The electron momentum is estimated by combining the energy measurement in the ECAL with the momentum measurement in the tracker. The momentum resolution for electrons with ${\pt \approx 45\GeV}$ from ${\PZ \to \Pe \Pe}$ decays ranges from 1.6 to 5.0\%. It is generally better in the barrel region than in the endcaps, and also depends on the bremsstrahlung energy emitted by the electron as it traverses the material in front of the ECAL~\cite{CMS:2020uim,CMS-DP-2020-021}.

Muons are detected in the region of ${\aeta<2.4}$ with drift tubes, cathode strip chambers, and resistive-plate chambers. Matching muons to tracks measured in the silicon tracker results in a \pt resolution of 1\% in the barrel and 3\% in the endcaps for muons with \pt up to {100\GeV}. Overall, the efficiency to reconstruct and identify muons is greater than 96\%~\cite{Sirunyan:2018fpa}. 

The electron (muon) isolation is determined relative to its ${\pt^{\ell}}$ values, where ${\ell}$ is \Pe (\PGm), by summing over the scalar \pt of the PF particles within a cone of ${\dr=\sqrt{\smash[b]{(\Delta\eta)^2+(\Delta\phi)^2}}=0.3}$ (0.4) around the lepton (where ${\phi}$ is azimuthal angle in radians), divided by ${\pt^{\ell}}$:
\ifthenelse{\boolean{cms@external}}{
  \begin{multline*}
    \label{eq:iso}
    \irel = \Bigg( \sum \pt^{\text{PV\,charged}} + \text{max}\Big[0, \sum \pt^{\text{neutral}} \\
          + \sum \pt^\Pgg - \pt^{\mathrm{PU}}(\ell) \Big] \Bigg) \Big/ \pt^\ell,
  \end{multline*}
  }{
  \begin{linenomath}
    \begin{equation}
    \label{eq:iso}
    \irel = \left( \sum \pt^{\text{PV\,charged}} + \text{max}\left[0, \sum \pt^{\text{neutral}} + \sum \pt^\Pgg - \pt^{\mathrm{PU}}(\ell)\right]\right) \biggm/ \pt^\ell,
    \end{equation}
  \end{linenomath}
}
where ${\pt^{\text{PV\,charged}}}$, ${\pt^{\text{neutral}}}$, and ${\pt^\Pgg}$ are the \pt of charged hadrons, neutral hadrons, and photons within the cone, respectively. The neutral particle contribution to the isolation from pileup, ${\pt^\mathrm{PU}(\ell)}$, is estimated for the electron from the area of jets and their median energy density in each event~\cite{Cacciari:2008gn}. For the muon, half of the \pt sum of the charged hadrons not coming from the PV within the isolation cone is used instead. The factor of 0.5 is estimated from simulations to be the ratio of neutral particle to charged hadron production in inelastic \Pp{}\Pp collisions~\cite{Sirunyan:2018fpa}. The charged-particle contribution to the isolation from pileup is rejected by requiring all tracks to originate from the PV. An isolation requirement of ${\irele<0.10}$ (${\irelm<0.15}$) is imposed to suppress backgrounds of jets misidentified as an electron (muon).

Charged hadrons are identified as charged particle tracks neither identified as electrons, nor as muons. Neutral hadrons are identified from HCAL energy clusters not assigned to any charged hadron or from an excess in the combined ECAL and HCAL energy with respect to the expected charged-hadron energy deposit.

For each event, hadronic jets are clustered from these reconstructed particles using the infrared and collinear safe anti-\kt algorithm~\cite{Cacciari:2008gp, Cacciari:2011ma} with a distance parameter of 0.4. Jet momentum is determined as the vectorial sum of all particle momenta in the jet, and is found from simulation to be, on average, within {5\%--10\%} of the true momentum over the whole \pt spectrum and detector acceptance. Pileup can contribute additional tracks and calorimetric energy depositions to the jet momentum. To mitigate this effect, charged particles identified to be originating from pileup vertices are discarded and an offset correction is applied to correct for remaining contributions. Jet energy corrections are derived from simulation to bring the measured response of jets to that of particle level jets on average. In situ measurements of the momentum balance in dijet, ${\text{photon}\text{+jet}}$, ${\PZ\text{+jet}}$, and multijet events are used to account for any residual differences in the jet energy scale between data and simulation~\cite{Khachatryan:2016kdb}. The jet energy resolution amounts typically to {15\%--20\%} at {30\GeV}, 10\% at {100\GeV}, and 5\% at {1\TeV}~\cite{Khachatryan:2016kdb}. Additional selection criteria are applied to each jet to remove jets potentially dominated by anomalous contributions from various subdetector components or reconstruction failures. Jets are required to have a ${\pt>30\GeV}$, ${\aeta<4.7}$, and be separated from each lepton of the identified \emu pair by ${\dr>0.4}$. Jets originating from {\PQb} hadron decays and detected within the tracker acceptance of ${\aeta<2.5}$ are tagged using a deep neural network based algorithm, \textsc{DeepJet}, using a working point with a 94\% b-jet identification efficiency at a 10\% misidentification rate for light-flavor quark and gluon jets in \ttbar events~\cite{Bols:2730134}.

Hadronic \PGt decays (\tauh) are reconstructed from jets, using the hadrons-plus-strips algorithm~\cite{Sirunyan:2018pgf}, which combines 1 or 3 tracks with energy deposits in the calorimeters, to identify the tau decay modes. To distinguish genuine \tauh decays from jets originating from the hadronization of quarks or gluons, and from electrons, or muons, the \textsc{DeepTau} algorithm is used~\cite{CMS:2022prd}. Information from all individual reconstructed particles near the \tauh axis is combined with properties of the \tauh candidate and the event. The rate at which jets are misidentified as \tauh by the \textsc{DeepTau} algorithm depends on the \pt and whether it was initiated by a quark or gluon. A working point with an 80\% \tauh identification efficiency and a {0.05\%--0.95\%} misidentification rate for jets is used.

The missing transverse momentum, \ptvecmiss, is computed as the negative vectorial \pt sum of all the PF candidates in an event~\cite{Sirunyan:2019kia}, with its magnitude labeled as \ptmiss. Corrections to the reconstructed jet energy scale are propagated to the \ptvecmiss. Anomalous high-\ptmiss events can originate from various reconstruction failures, detector malfunctions, or noncollision backgrounds. Such events are rejected using event filters designed to identify more than {85\%--90\%} of the spurious high-\ptmiss events with a misidentification rate of less than 0.1\%~\cite{Sirunyan:2019kia}. 

\section{Event selection}
\label{sec:selection}

The signal topology consists of an oppositely charged electron-muon pair with possible additional jets. Events with an oppositely charged electron-muon pair separated by ${\dr>0.3}$ are selected. Both the electron and muon are required to have a longitudinal and a transverse impact parameter within 5 and {2\unit{mm}} from the PV, respectively.

The invariant mass of the \emu pair, \mem, is required to fall in the range of {100--170\GeV} such that signals with the lowest (highest) ${\mx=110\ (160)\GeV}$ targeted in this search are fully contained. The \mem window is intentionally chosen to lie beyond the peak of the \ttbar background distribution, thus selecting a region where it falls smoothly. Backgrounds from ${\PH\to\PGt\PGt}$ and ${\PH\to\PW\PW}$ also peak below the mass window since part of the \PH four-momentum is carried away by the final-state neutrinos. 

The \pt of the electron (muon), ${\pt^{\Pe}}$ (${\pt^{\PGm}}$), collected by the single-electron (single-muon) triggers is required to be larger than {29 (26)\GeV} in 2016, {34 (29)\GeV} in 2017, and {34 (26)\GeV} in 2018. These \pt requirements are chosen to be slightly above the \pt thresholds of the triggers so that the efficiency of the triggers is nearly 100\%. For electrons (muons) that do not pass the single-electron (single-muon) trigger requirements, their \pt are required to be larger than {25 (20)\GeV} in all years. Events containing additional reconstructed electrons, muons, or hadronically decaying tau candidates are vetoed. Events with at least one {\PQb}-tagged jet are also vetoed to suppress the \ttbar and single top quark backgrounds.  

\section{Event categorization}
\label{sec:categorization}

\begin{figure}[!htb]
  \centering
  \includegraphics[width=0.48\textwidth]{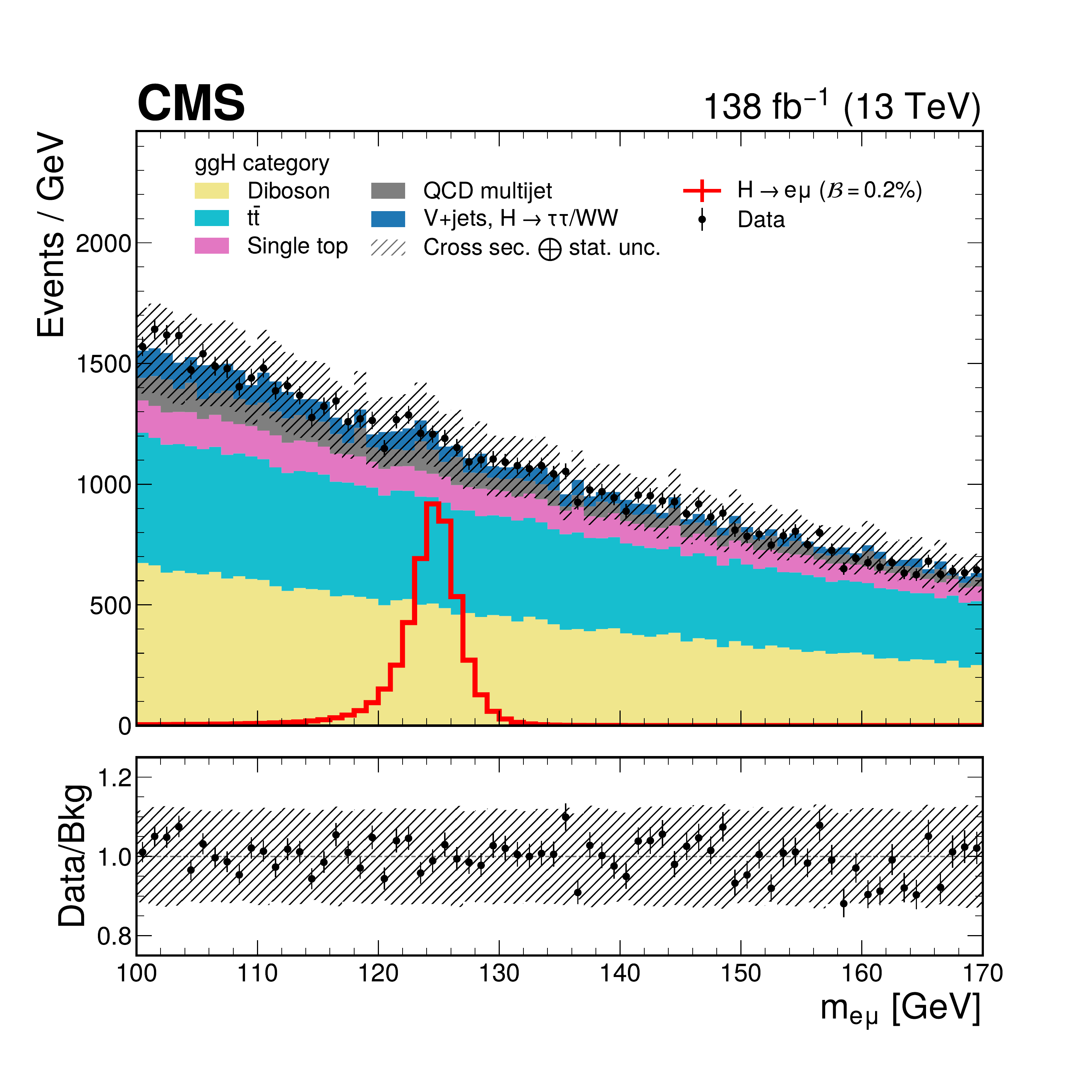}
  \includegraphics[width=0.48\textwidth]{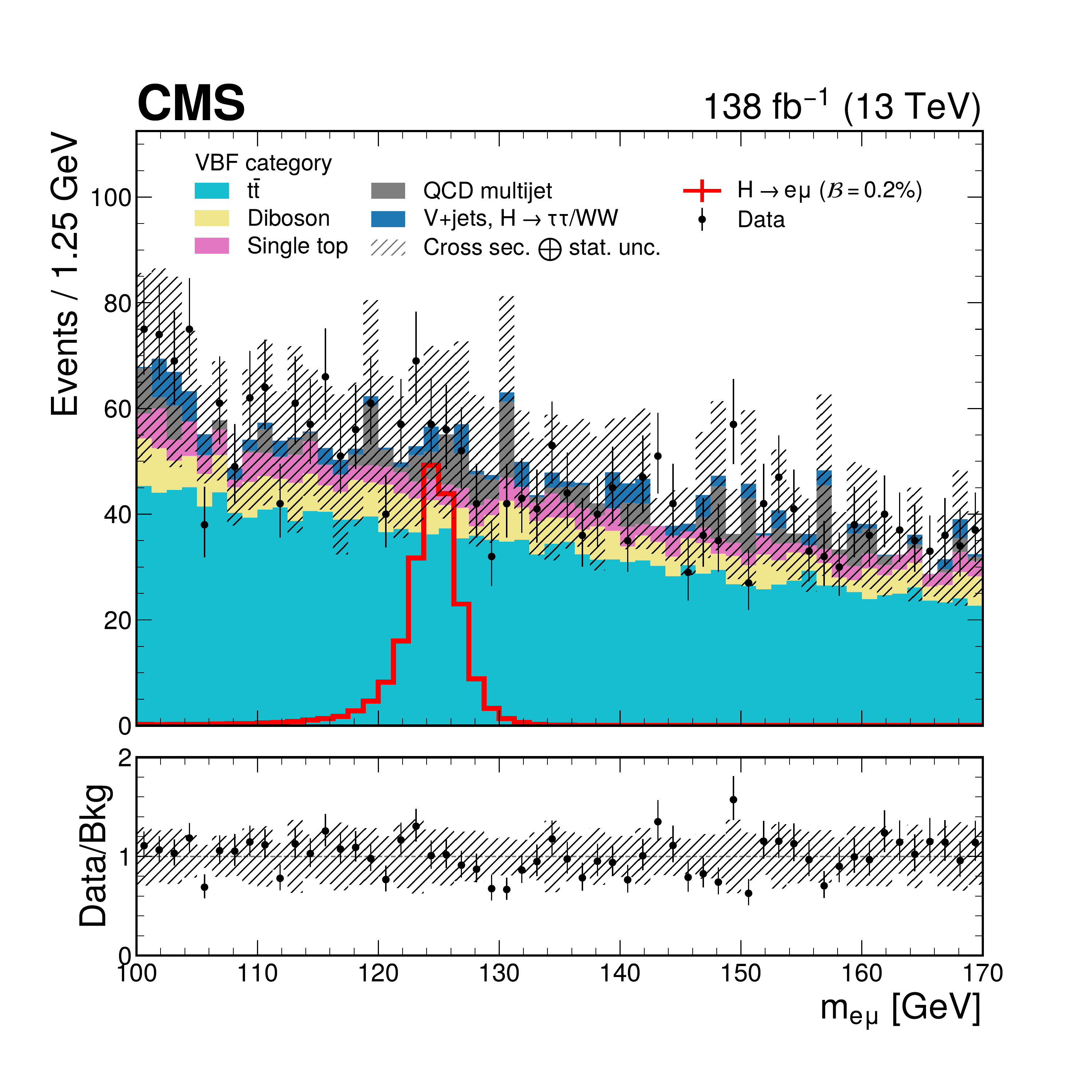}
  \caption{The \mem distributions of the data, simulated backgrounds and signals of \HemSM in the \ggH (\cmsLeft) and the VBF categories (\cmsRight). A ${\BHemSM=0.2\%}$ is assumed for the signal for illustration. The lower panel in each plot shows the ratio of the data to the total estimated background. The uncertainty band corresponds to the background uncertainties, adding in quadrature the statistical and the SM cross section uncertainties.}
  \label{fig:mem_distribution} 
\end{figure}

Events are first divided into two broad categories to enhance the signal from either the \ggH or the VBF production mechanisms. Events with two or more jets where the two highest \pt jets have an invariant mass ${\mjj >400\GeV}$ and a pseudorapidity separation ${\detajj > 2.5}$ are classified as the VBF production category. Otherwise, events enter the \ggH production category. The \mem distributions of the data, the simulated backgrounds, and signals of \HemSM are shown in Fig.~\ref{fig:mem_distribution} for both categories. The QCD multijet background shown is estimated from a control region of the data using events with an \emu pair of the same electric charge and extrapolated to the signal region as a function of jet multiplicity and the \dr separation of the \emu pair as described in Ref.~\cite{Sirunyan:2766068}. The data and background simulations show good agreement within the statistical and the SM cross section uncertainties combined.

\begin{figure}[!htb]
  \centering
  \includegraphics[width=0.48\textwidth]{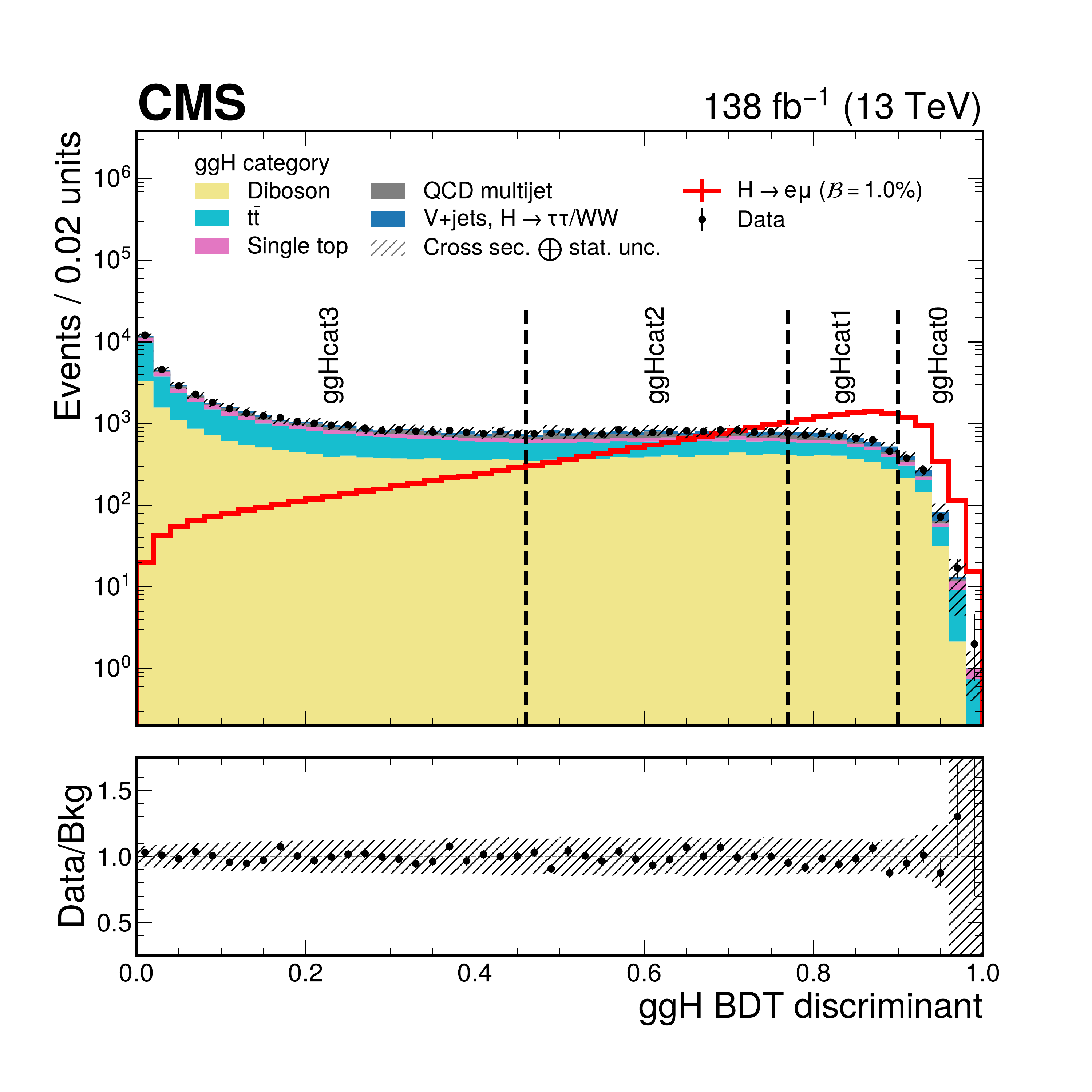}
  \includegraphics[width=0.48\textwidth]{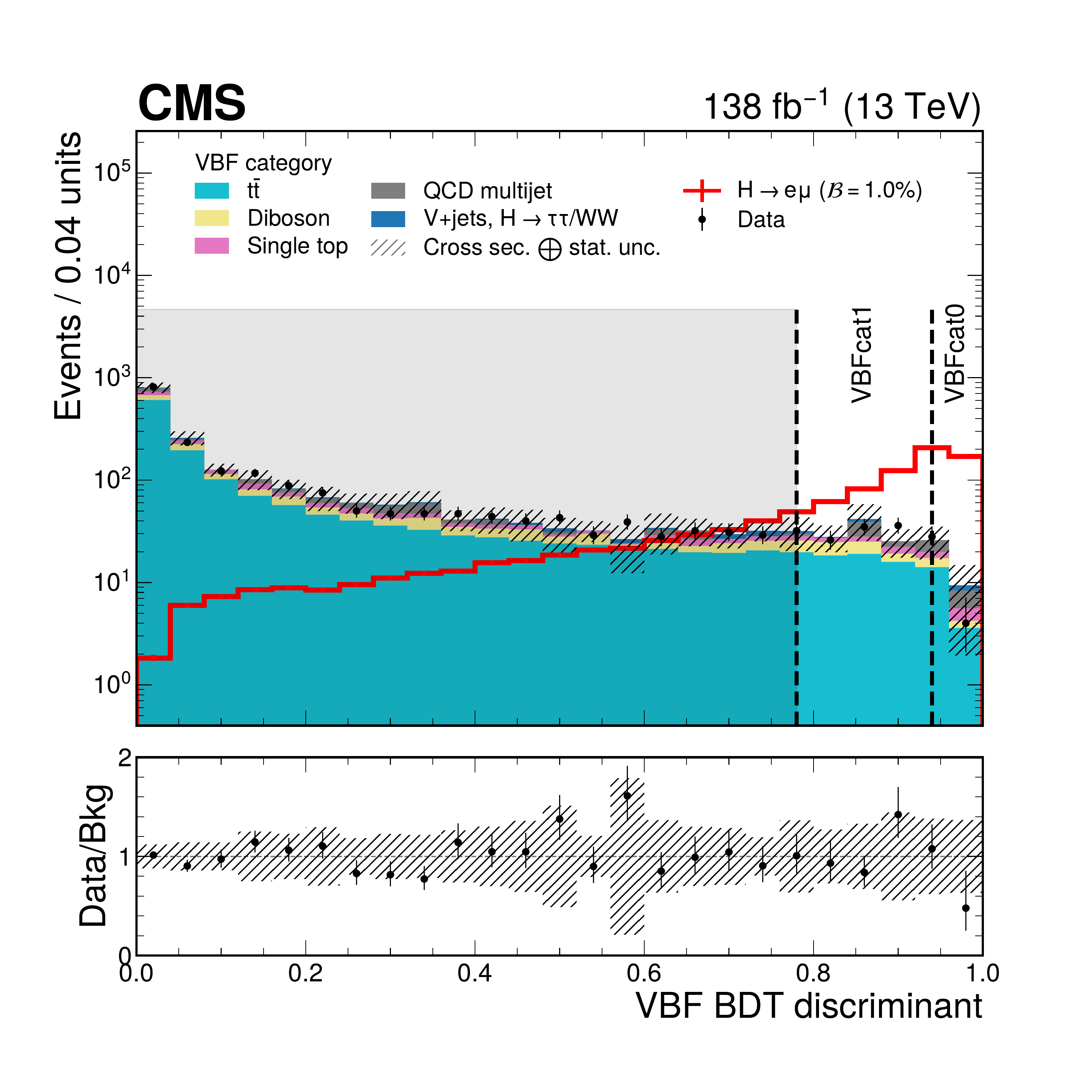}
	\caption{The \ggH and VBF BDT discriminant distributions of the data, simulated backgrounds and signals of \HemSM for each BDT trained in the \ggH (\cmsLeft) and the VBF categories (\cmsRight). A ${\BHemSM=1.0\%}$ is assumed for the signal for illustration. The lower panel in each plot shows the ratio of the data to the total estimated background. The uncertainty band corresponds to the background uncertainties, adding in quadrature the statistical and the SM cross section uncertainties. Vertical lines in the plots illustrate boundaries of the subcategories: {\ggH cat 0--3} and {VBF cat 0--1}, as defined in Sec.~\ref{sec:sensitivity}. Events in the shaded region of the VBF category with a VBF BDT discriminant less than 0.78 are discarded since their sensitivity is an order of magnitude lower than other subcategories.}
  \label{fig:bdt_distribution} 
\end{figure}

The two broad categories are further split according to the signal purity using the output of boosted decision trees (BDTs) trained with the \textsc{XGBoost} package~\cite{DBLP:journals/corr/ChenG16}. The BDTs are trained separately for the \ggH and the VBF categories. The BDT discriminants range from 0 for backgroundlike events to 1 for signal-like events. For both BDTs, a mixture of simulated signal events is used in the training including events of \HemSM and \Xem at ${\mx=110}$, 120, 130, 140, 150, and {160\GeV} from both the \ggH and the VBF production modes. Kinematics variables from the dominant sources of backgrounds of di-leptonic decays of \ttbar and WW diboson events are used in the training. All events used in the training are from Monte Carlo (MC) simulations described in Sec.~\ref{sec:dataset}. The simulated signals of \HemSM and backgrounds are weighted according to their expected yields from the SM cross sections. The simulated signal samples of \Xem are weighted according to their relative SM-like production cross sections as evaluated in Ref.~\cite{deFlorian:2016spz} as a function of \mx. Their total weights are matched to that of the backgrounds in the training to ensure the larger total weights of the background samples does not lead to BDTs with poor signal identification efficiency. Each signal event is additionally reweighted by the inverse of its expected mass resolution during training. The mass resolution is the uncertainties of \mem propagated from the expected uncertainty of the lepton \pt measurements. This reweighting allows the BDTs to assign more importance in classifying signal events with high mass resolution. The \ggH and VBF BDT discriminant distributions of the data, the simulated backgrounds, and signals of \HemSM are shown in Fig.~\ref{fig:bdt_distribution} for the \ggH and VBF category, respectively. The data and background simulations show good agreement within the statistical and the SM cross section uncertainties combined.

\subsection{BDT input variables}
\begin{figure}[!htb]
\centering
  \includegraphics[width=0.48\textwidth]{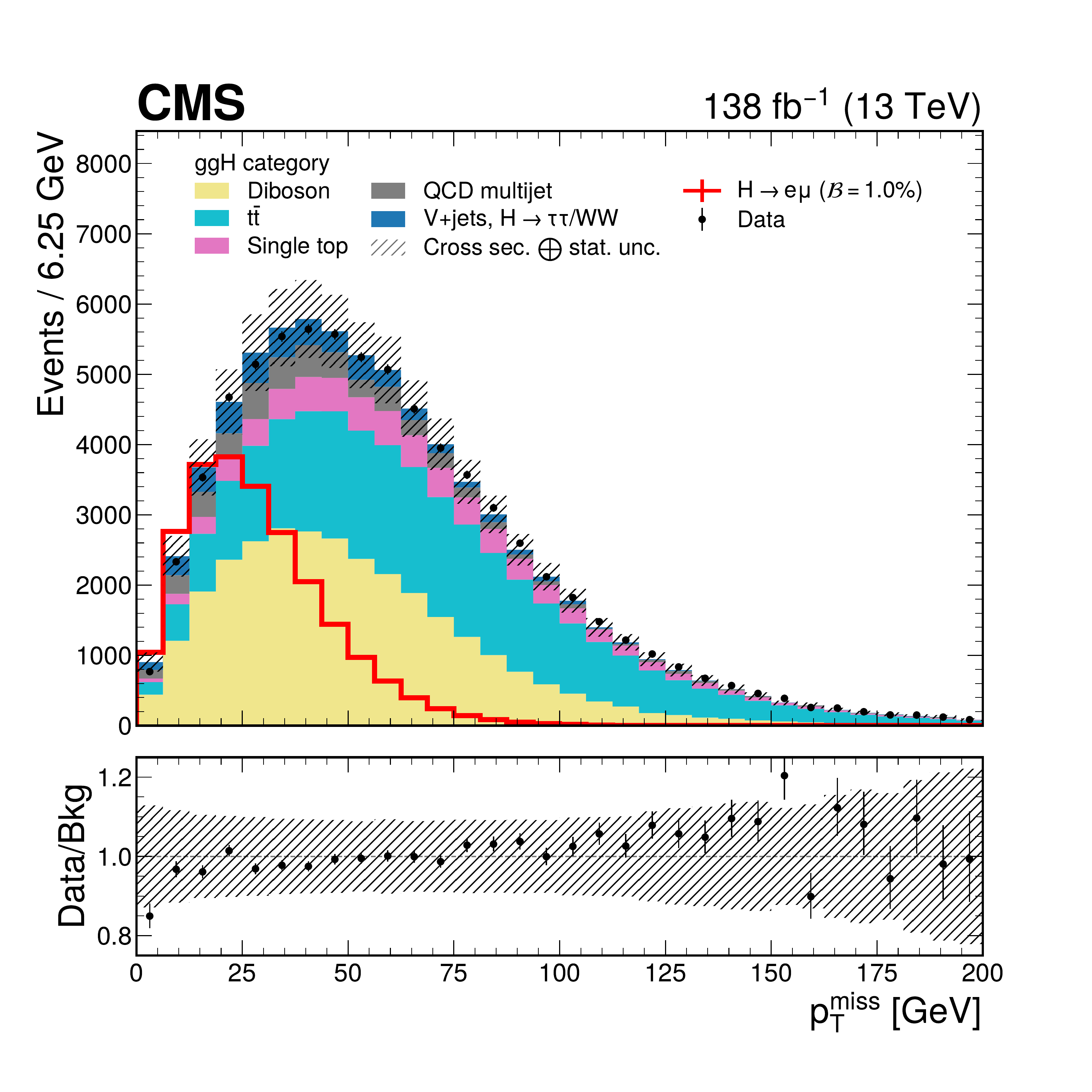}
  \includegraphics[width=0.48\textwidth]{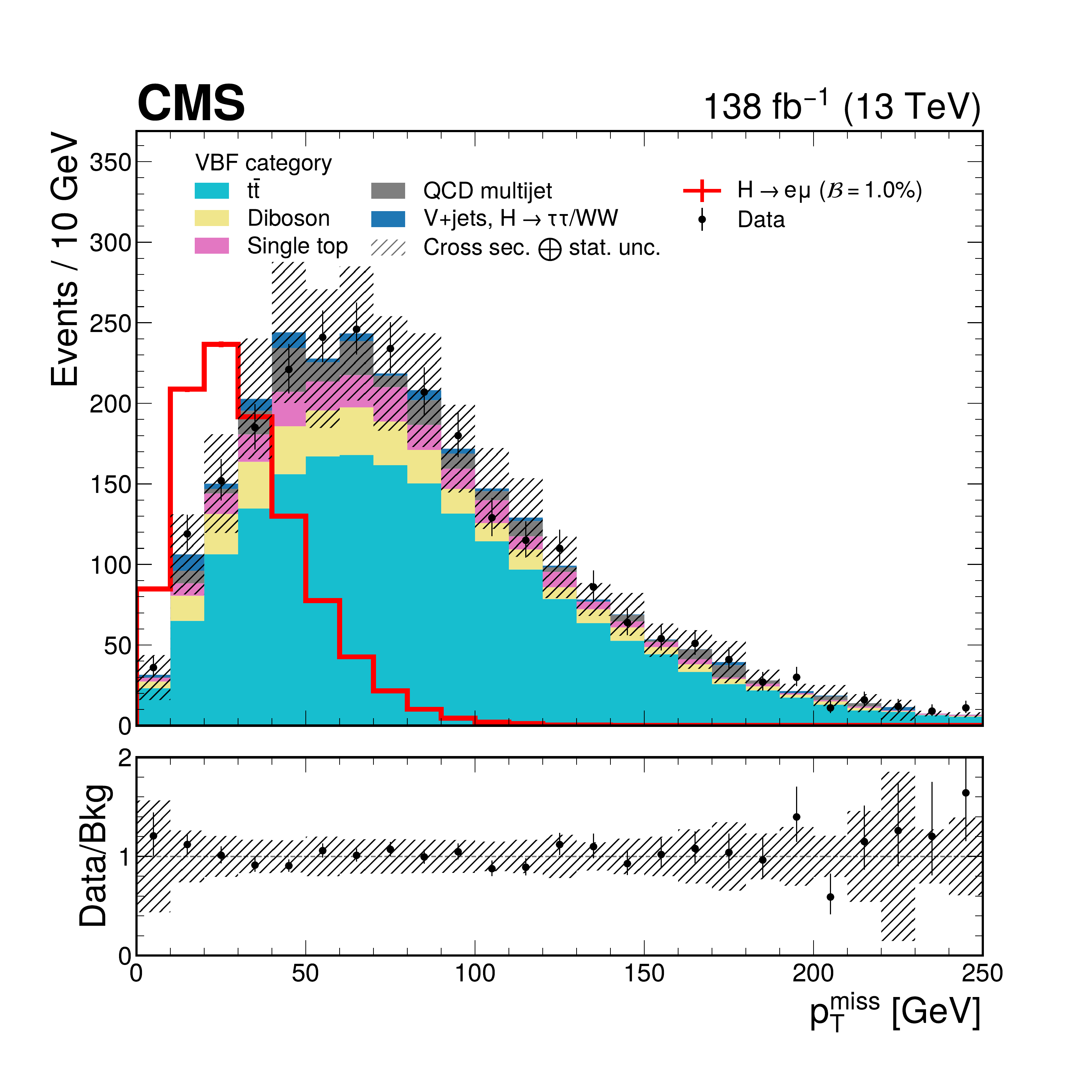}
  \caption{The \ptmiss distributions of the data, simulated backgrounds and signals of \HemSM in the \ggH (\cmsLeft) and the VBF categories (\cmsRight). A ${\BHemSM=1.0\%}$ is assumed for the signal for illustration. The lower panel in each plot shows the ratio of the data to the total estimated background. The uncertainty band corresponds to the background uncertainties, adding in quadrature the statistical and the SM cross section uncertainties.}
  \label{fig:met_distribution}
\end{figure}

The BDT input variables are chosen such that the BDTs do not make use of \mem to discriminate between the signal and background. This ensures background events with \mem close to the signal resonance are not preferrentially assigned a higher BDT discriminant which distorts their smoothly falling shape to form spurious signal resonances along \mem. For example, the \emu system's \pt scaled by \mem, ${\pt^{\Pe\PGm}/\mem}$, is used instead of ${\pt^{\Pe\PGm}}$ which is correlated with \mem. The background samples are also reweighted to match the shape of the \mem distribution of the signals during training to further ensure that the BDTs do not benefit from using \mem to discriminate between the signal and background. No sculpting of the \mem distribution is observed for the MC background events in different ranges of the BDT discriminants. The BDT discriminant distributions of the simulated signals at different Higgs boson masses are also observed to be similar. The distributions of \ptmiss, which is the most discriminating variable in both the \ggH and VBF category, are shown in Fig.~\ref{fig:met_distribution} for both categories.

\subsubsection{The \texorpdfstring{${\ggH}$}{ggH} BDT}
The input variables to the \ggH BDT include the absolute pseudorapidities of the electron, ${\abs{\eta^{\Pe}}}$, and of the muon, ${\abs{\eta^{\PGm}}}$, the ratio of the \emu system's \pt to its invariant mass, ${\pt^{\Pe\PGm}/\mem}$, and the pseudorapidity separation of the \emu pair, ${\abs{\Delta\eta(\Pe,\PGm)}}$. Variables related to the \ptvecmiss including \ptmiss and its azimuthal separation to the \emu system, ${\Delta\phi(\ptvecmiss,\pt^{\Pe\PGm})}$, are also used to discriminate the neutrinoless LFV decay against backgrounds with neutrinos in the final state. The number of jets in each event is also added as an input variable.

Additional jet variables are added for events with at least one jet, including the \pt of the leading jet, ${\pt^{j_{1}}}$, the absolute pseudorapidity of the leading jet, ${\abs{\eta^{j_{1}}}}$, and the pseudorapidity separation of the leading jet to the \emu system, ${\abs{\Delta\eta(j_{1},\Pe\PGm)}}$. For events with at least two jets, the scalar \pt of all jets is added. Observables sensitive to the angular and \pt correlations between the \emu system and the two highest \pt jets are also included, including the \pt-balance ratio: 
\begin{linenomath}
\begin{equation}\label{eqn:Rpt}
	\pt\text{-balance ratio} = \frac{\abs{\vec{\pt}^{\Pe}+\vec{\pt}^{\PGm}+\vec{\pt}^{j_1}+\vec{\pt}^{j_2}}}{\pt^{\Pe\PGm}+p^{j_1}_T+\pt^{j_2}},
\end{equation}
\end{linenomath}
and the ${\pt}$-centrality:
\begin{linenomath}
  \begin{equation}
    \pt\text{-centrality} = \frac{\pt^{\Pe\PGm}-(\pt^{j_{1}}+\pt^{j_{2}})/2}{\pt^{j_{1}}-\pt^{j_{2}}}.
  \end{equation}
\end{linenomath}

If jets are absent in an event, the undefined jet variables are handled by the sparsity-aware split finding algorithm in the \textsc{XGBoost} package~\cite{DBLP:journals/corr/ChenG16}, with the exception of ${\pt^{j_{1}}}$ set to be zero in events with no jets. When jet variables are used at a decision split of a tree, the sparsity-aware algorithm assigns events with an undefined value to the direction that minimizes the loss function. 

\subsubsection{The VBF BDT}
The input variables to the VBF BDT are the same to that of the \ggH BDT with a few exceptions: ${\Delta\phi(\ptvecmiss,\pt^{\Pe\PGm})}$, ${\abs{\eta^{j_{1}}}}$, and ${\abs{\Delta\eta(j_{1},\Pe\PGm)}}$ are dropped due to their insignificant contributions to the VBF BDT training. Instead, the Zeppenfeld variable~\cite{2013Z}, defined as
\begin{linenomath}
  \begin{equation}
    \text{Zeppenfeld variable} = \frac{\eta^{\Pe\PGm}-(\eta^{j_{1}}+\eta^{j_{2}})/2}{\detajj},
  \end{equation}
\end{linenomath}
is added along with \mjj and \detajj.
\subsection{Sensitivity optimization}
\label{sec:sensitivity}
The \ggH and VBF categories are further split according to the \ggH and VBF BDT discriminant value to optimize the expected sensitivity of the search. The expected sensitivity is estimated from the expected significance of discovery in the asymptotic approximation~\cite{Cowan:2010js} from a signal-plus-background (${S+B}$) fit to the \mem distribution in the data within {100--170\GeV}, overlayed with a simulated signal of ${\BHemSM = 5.9\times10^{-5}}$, the most stringent direct expected limit of \BHemSM up-to-date~\cite{Aad:2690381}. In these fits, the signal peaks are modeled with a sum of three (two) Gaussians for the \ggH (VBF) production signals in the \ggH categories. For both the \ggH and VBF production signals in the VBF categories, a sum of two Gaussians is used. The number of Gaussians chosen are motivated by the likelihood-ratio test~\cite{Wilks:1938dza}, as explained in Sec.~\ref{sec:modeling}. The total expected background is modeled from data directly with a third (first) order Bernstein polynomial for the \ggH (VBF) category. Subcategory boundaries are determined separately in the \ggH and the VBF categories by iteratively scanning in steps of 0.01 for a cutoff along the \ggH and the VBF BDT discriminants, respectively, that maximizes the total expected sensitivity. This procedure is repeated until the further gain in sensitivity is less than 1\%.

Four optimized subcategories are defined for the \ggH category, named as ``ggH cat 0'', ``ggH cat 1'', ``ggH cat 2'', and ``ggH cat 3'', which correspond to events of decreasing ggH BDT discriminant of {0.89--1.00}, {0.77--0.89}, {0.46--0.77}, and {0.00--0.46}, respectively. Similarly, three optimized subcategories are defined for the VBF category, ``VBF cat 0'', ``VBF cat 1'', and ``VBF cat 2'', corresponding to events with a VBF BDT discriminant between {0.94--1.00}, {0.78--0.94}, and {0.00--0.78} respectively. Events from the least sensitive category ``VBF cat 2'' are discarded. Table~\ref{tab:order} summarizes the definition, the expected background (${B}$), and signal yield of \HemSM at ${\mathcal{B}=10^{-4}}$ (${S}$) in each categories at an integrated luminosity of {138\fbinv}. An estimate of the expected significance in each category by ${S/\sqrt{B}}$ is also listed. The yields are estimated by the number of MC events within a \mem interval of ${125\GeV\pm\sigma_\text{eff}}$, where ${\sigma_\text{eff}}$ is half of the smallest symmetric interval that contains 68\% of the signal events in each category.

\begin{table*}[!htb]
\centering
\caption{Range of the \ggH (VBF) BDT discriminant to define the ggH (VBF) subcategories, and the corresponding expected background (${B}$), and signal yield of \HemSM at ${\mathcal{B}=10^{-4}}$ (${S}$) at an integrated luminosity of {138\fbinv}. The yields are estimated by the number of MC events within a \mem interval of ${125\GeV\pm\sigma_\text{eff}}$, where ${\sigma_\text{eff}}$ is half of the smallest symmetric interval that contains 68\% of the signal events in each category. The fraction contributions of the expected signal yields from the \ggH and VBF production mode are listed. An estimate of the expected significance in each category by ${S/\sqrt{B}}$ is also listed.}
\cmsTable{
\begin{scotch}{lccccccc}
Analysis    & BDT& ${\sigma_\text{eff}}$ & \multirow{2}{*}{${S}$}& \ggH mode & VBF mode & \multirow{2}{*}{${B}$} & \multirow{2}{*}{${S/\sqrt{B}}$} \\
	category  & discriminant & (\GeVns)& & fraction (\%) & fraction (\%) & & \\
\hline
ggH cat 0 & 0.89--1.00 & 1.7 & 22.4 & 94.2 & 5.8  & 79.1 & 2.5    \\
ggH cat 1 & 0.77--0.89 & 2.1 & 55.4 & 96.4 & 3.6  & 399.3 & 2.8   \\
ggH cat 2 & 0.46--0.77 & 2.4 & 60.4 & 96.0 & 4.0  & 1045.9 & 1.9  \\
ggH cat 3 & 0.00--0.46 & 2.5 & 20.9 & 94.4 & 5.6  & 3755.4 & 0.3  \\
VBF cat 0 & 0.94--1.00 & 1.9 & 2.2  & 23.7 & 76.3 & 1.1 & 2.2     \\
VBF cat 1 & 0.78--0.94 & 2.2 & 2.4  & 42.2 & 57.8 & 9.7 & 0.8    \\
VBF cat 2 & 0.00--0.78 & 2.4 & 2.3  & 61.8 & 38.2 & 161.3 & 0.2   \\
\end{scotch}
\label{tab:order}
}
\end{table*}

\section{Signal and background modeling}
\label{sec:modeling}
\begin{figure}[!htb]
  \centering
  \includegraphics[width=0.48\textwidth]{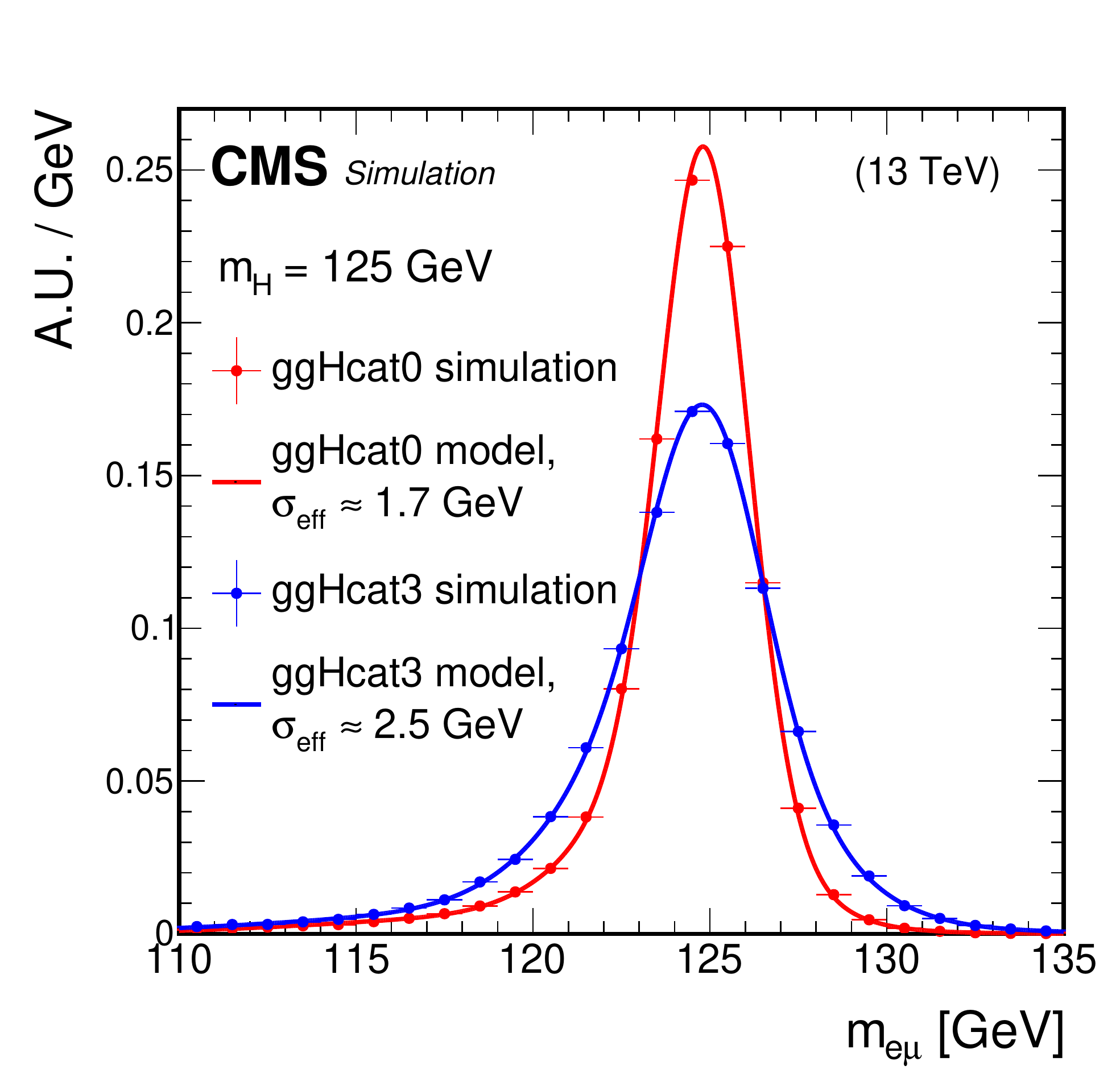}
  \includegraphics[width=0.48\textwidth]{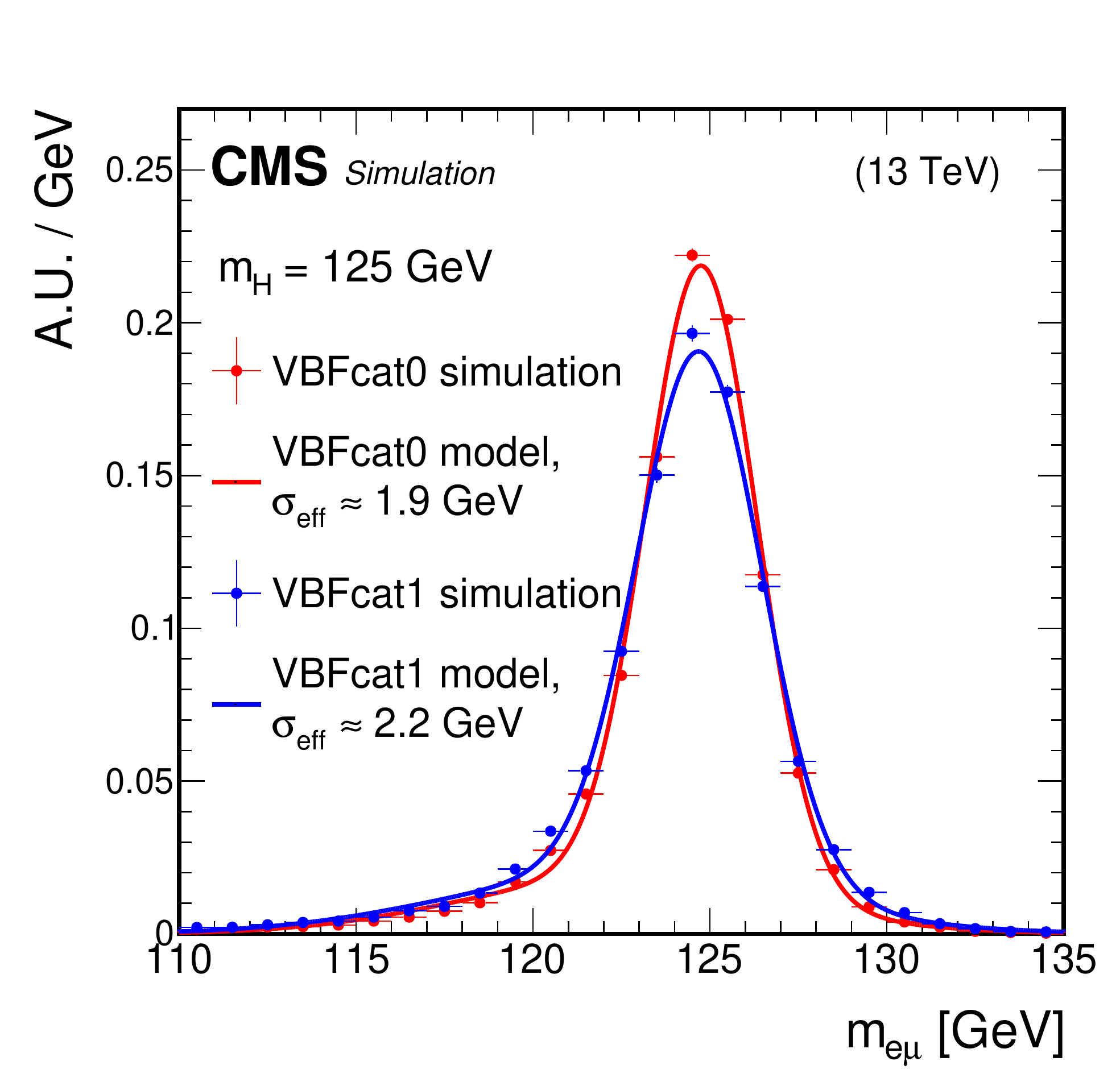}
  \caption{Example fits of the signal models to the simulated ${\HemSM}$ signal in the analysis categories ggH cat 0 and ggH cat 3  (\cmsLeft), as well as VBF cat 0 and VBF cat 1 (\cmsRight), summing events from both the \ggH and VBF production modes. Half of the smallest symmetric \mem interval that contains 68\% of the signal events, ${\sigma_{\text{eff}}}$, is shown in the legends for each signal as an illustration of the signal resolution. The signal resolution in general improves with the signal purity of the analysis categories.}
  \label{fig:signal_fit} 
\end{figure}

The \mem distributions of simulated signal events are fit with a sum of Gaussian distributions for each production mode, category, and mass of the Higgs boson. The number of Gaussians is chosen with the likelihood ratio test~\cite{Wilks:1938dza}, such that the next higher order does not give a significantly better fit at a $p$-value of 0.05. A sum of three Gaussians is determined to be sufficient for the signals from the \ggH production mode in the \ggH category, while a sum of two Gaussians is sufficient for the rest. When carrying out the fits, the means are fit as a sum of the known simulated \mx or \mh and a small floating shift due to initial/final-state radiations and detector effects. Example fits of the signal models to the simulated ${\HemSM}$ signal are shown in Fig.~\ref{fig:signal_fit} for the analysis categories ggH cat 0 and ggH cat 3, as well as VBF cat 0 and VBF cat 1, summing events from both the \ggH and VBF production modes. ${\sigma_{\text{eff}}}$ for each distribution is included as an illustration of the signal resolution. The signal resolution in general improves with the signal purity of the analysis categories since signal events are reweighted by the inverse of their mass resolution during training of the BDTs as mentioned in Sec.~\ref{sec:categorization}. The \mem distributions of a Higgs boson with mass between the simulated mass points are interpolated by fitting the parameters and normalizations of the sum of Gaussians with second-order polynomials as a function of the Higgs boson mass.

The background in each category is modeled with a Bernstein polynomial. Orders of the polynomials are chosen with a bias study as follows. The \mem distribution in data from {100--170\GeV} is first fit with three distinct functional forms: a Bernstein polynomial, a sum of exponential functions, and a sum of power law functions. An optimal order for each function is chosen with the likelihood-ratio test~\cite{Wilks:1938dza} at a $p$-value of 0.05. Then, ensembles of 2000 pseudoexperiments are generated with the \mem distributions drawn from each of the three background models, with or without an injection of a signal at the simulated \mx points with a branching fraction of ${10^{-4}}$. The pseudoexperiment mass spectra are fit with a Bernstein polynomial with an order equal to or higher than the chosen order in the first step. The signal yield from these fits would in general differ from the injected yield since different background models are used to generate and fit the signal peaks. The bias of a model choice is evaluated as the average difference of the fit signal yield to the generated yield divided by the uncertainty in the fit yield in the pseudoexperiments. The final order of the Bernstein polynomial used to model the background in each category is then chosen by requiring the bias to be less than 20\% across all generating functions and ensembles of pseudoexperiments. The third order is chosen for all \ggH subcategories, while the second and the first order are chosen for ``VBF cat 0'' and ``VBF cat 1'', respectively.

\section{Systematic uncertainties}
\label{sec:sysUnc}
\subsection{Background uncertainties}
The systematic uncertainty associated to the bias of the background model choice is modeled by adding a signal-like background shape to the background models. The signal-like background shape is drawn directly from the signal models in each category. The normalization of the signal-like background is implemented as a nuisance parameter modeled with a Gaussian constraint of zero mean and a standard deviation equal to the maximum of the pseudoexperiment averaged signal yield fit over the three background models in the bias study with no signal injected, as described in Sec.~\ref{sec:modeling}. The maxima are no more than 20\% of the statistical uncertainties in the fits. The standard deviations amount to a Higgs-like signal yield with a branching fraction ${\mathcal{B}(\PH/\PX\to\Pe\PGm)}$ of {0.4--${2.9\times10^{-5}}$} across the categories. This is a dominant source of systematic uncertainty, contributing {6.9--14.4\%} of the total uncertainty on the best fit of the signal yield, depending on {\mh/\mx}. Besides the systematic uncertainty associated to the bias of the background model and the statistical uncertainty of the fits, there are no additional systematic uncertainties in the background models as they are derived directly from data.

\subsection{Signal uncertainties}
The simulated signals are affected by various sources of experimental and theoretical systematic uncertainties. These uncertainties affect both the yield and the shape of the \mem distributions. The systematic uncertainties are incorporated as nuisance parameters in the ${S+B}$ likelihood fit of the \mem distribution. Log-normal constraints are assumed for uncertainties affecting the yield, and Gaussian constraints are assumed for uncertainties affecting the fit parameters of the \mem distribution. The uncertainties affecting the yield have negligible effects on the signal shapes in general. All the uncertainties are treated as correlated between the categories, except for systematic uncertainties from the interpolation of signal shapes. The list of yield uncertainties is summarized in Table~\ref{tab:systematics} for the \ggH and VBF production modes separately.

\begin{table*}[!htb]
\centering
\caption{Systematic uncertainties in the expected signal yields from different sources for the \ggH and VBF production modes. All the uncertainties are treated as correlated among categories. The ranges listed are for signals with a different Higgs boson mass.}
\begin{scotch}{lcc}
Systematic uncertainties                                & \ggH mode (\%)  & VBF mode (\%)\\
\hline
Muon identification, isolation, and trigger             & 0.2--0.4     & 0.3--0.4      \\
Electron identification, isolation, and trigger         & 1.8--2.6     & 2.0--2.5      \\
\PQb-tagging veto efficiency                           & 0.1--0.4     & 0.1--0.3      \\
Jet energy scale                                        & 0.6--18.6    & 4.0--10.0     \\
Unclustered energy scale                                & 0.1--9.3     & 0.1--9.9      \\
Trigger timing inefficiency                             & 0.1--0.4     & 0.2--0.5      \\
Integrated luminosity                                   & 1.6          & 1.6          \\
Pileup                                                  & 0.1--1.6     & 0.1--1.6      \\
Parton shower                                           & \NA          & 1.9--11.4     \\
Renormalization and factorization scales                & 3.9--8.0     & 0.2--0.5      \\
PDF+\alpS                                               & 3.0--3.2     & 1.9--2.1      \\
Effect of the ren. and fact. scales on the acceptance   & 0.2--11.2    & 0.2--1.3      \\
Effect of the PDF+\alpS on the acceptance               & 0.1--0.6     & 0.1          \\
Total                                                   & 2.9--23.8    & 5.2--16.3   \\
\end{scotch}
\label{tab:systematics}
\end{table*}

\subsubsection{Signal shape uncertainties}

The uncertainties in the electron (muon) momentum scale and resolution affect the means and widths of the signal models. These uncertainties are measured in \Zee (\Zmm) events in data and simulation in the ${\PH\to\PZ\PZ\to 4\ell}$ ${(\ell=\Pe,\PGm)}$ analysis~\cite{CMS:2021ugl}. They are estimated to be 0.1\% for the means and 10.0\% for the widths of the signal models.

\subsubsection{Signal yield uncertainties}
The uncertainties in the reconstruction, single-lepton trigger, offline identification, and isolation efficiencies of electrons and muons are respectively measured in \Zee and \Zmm events with the ``tag-and-probe'' method~\cite{Khachatryan:2010xn} in data and simulated events. They amount to be {1.8--2.6\%} for electrons and {0.2--0.4\%} for muons~\cite{CMS:2020uim, Sirunyan:2018fpa}. The lepton identification and isolation uncertainties are treated as correlated between the data-taking years, while the trigger uncertainties are treated as uncorrelated.

The uncertainties in the jet energy scale and resolution from different sources are evaluated as functions of the jet \pt and ${\eta}$~\cite{Khachatryan:2016kdb}. Jets with ${\pt<10\GeV}$ are classified as unclustered energy. The uncertainties in the unclustered energy scale for charged particles, neutral hadrons, photons, and very forward particles are evaluated separately according to the resolution of the different sub-detectors. The combined uncertainty of the unclustered energy scale is then propagated to the \ptvecmiss. Uncertainties on jets and \ptvecmiss affect both the \ggH and VBF BDTs, which are used to define the categories. They are transformed into signal yield uncertainties per category, which in turn enter as nuisance parameters in the likelihood fit. The efficiency to identify a {\PQb}-tagged jet with the \textsc{DeepJet} algorithm is different in data and simulations and affect the {\PQb}-tagging veto. Scale factors dependent on the jet \pt and ${\eta}$ are applied to correct the simulation~\cite{Sirunyan:2017ezt, CMS-DP-2018-033}. The uncertainties in these scale factors are taken into account.

The theoretical uncertainties in the renormalization and factorization scales, the choice of PDFs, and the value of the strong coupling constant, \alpS, evaluated at the \PZ boson mass, affect the measurement of the Higgs boson production cross sections~\cite{deFlorian:2016spz}. These uncertainties in turn affect the expected signal yield and are treated as correlated between the data-taking years. The QCD scales variations lead to {3.9--8.0\%} and {0.2--0.5\%} of uncertainty in the \ggH and VBF cross sections, respectively, while changes in the PDFs and \alpS result in {3.0--3.2\%} and {1.9--2.1\%} uncertainties, respectively. The uncertainties in the event acceptance in each category due to the scales, PDFs, and \alpS are also taken into account. An additional uncertainty in the VBF parton shower model is assigned as the signal yield difference between the dipole shower in \PYTHIA and the alternative angular-ordered shower in \HERWIG. This amounts to {1.9--11.4\%} uncertainties across the categories.

The integrated luminosities for the 2016, 2017, and 2018 data-taking years have {1.2--2.5\%} individual uncertainties~\cite{CMSlumi2016, CMS-PAS-LUM-17-004, CMS-PAS-LUM-18-002}, while the overall uncertainty for the {2016--2018} period is 1.6\%. They affect the overall yield of the signal expected from simulations. The uncertainty on the number of pileup vertices is evaluated by varying the pileup correction weights applied to the simulation. The variation of weight is obtained through a ${\pm4.6}$\% change to the total inelastic cross section at a nominal value of {69.2\mb} used to estimate the pileup effect on data. The pileup uncertainties are treated as correlated between the years.

During the 2016 and 2017 data-taking periods, a gradual timing shift of the signals from the ECAL first-level trigger caused a specific trigger inefficiency in the region of ${\aeta>2.0}$. For events containing an electron with ${\pt>50\GeV}$ or a jet with ${\pt>100\GeV}$, in the region of ${2.5<\aeta<3.0}$, the efficiency loss is between {10.0--20.0\%}, dependent on \pt, ${\eta}$, and time. Scale factors are computed to correct the detector acceptance in simulations to reflect this effect in the data. The uncertainty due to this correction is {0.1--0.5\%} and is treated as correlated between the two years.

\section{Results}
\label{sec:results}
\subsection{Results for the Higgs boson}
\begin{figure}[!htb]
  \centering
  \includegraphics[width=0.48\textwidth]{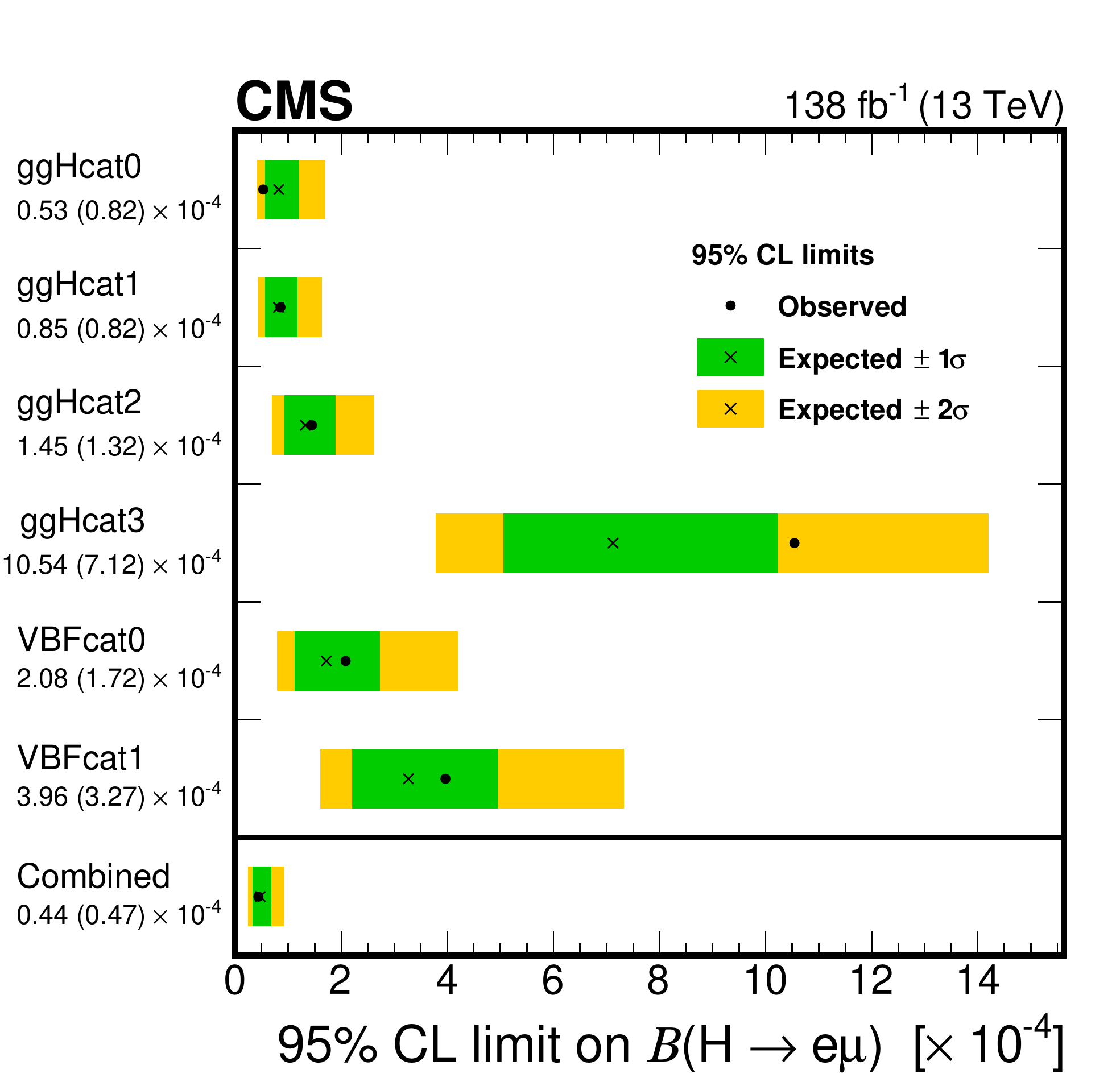}
  \caption{Observed (expected) 95\% \CL upper limits on \BHemSM for each individual analysis category (as shown in the left axis label) and for the combination of all analysis categories.}
  \label{fig:limit_emu_125} 
\end{figure}

No excess of data above the background prediction has been observed for the LFV decay of \HemSM. An upper limit on the branching fraction of the decay is computed using the \CLs criterion, with the profile likelihood as the test statistic, which is assumed to be distributed as in the asymptotic approximation~\cite{Junk:1999kv, Read:2002hq, Cowan:2010js}. The observed (expected) upper limit on \BHemSM is ${4.4\ (4.7)\times10^{-5}}$ at 95\% \CL. A breakdown of the upper limit on \BHemSM is shown per analysis category, and for the combination of all analysis categories is illustrated graphically in Fig.~\ref{fig:limit_emu_125} and listed in Table~\ref{tab:limit_emu_125}. Tabulated results are provided in the HEPData record~\cite{hepdata}.

\begin{table*}[!htb]
\centering
\caption{Observed and expected 95\% \CL upper limits on \BHemSM for each individual analysis category and for the combination of all analysis categories.}
\cmsTable{
\begin{scotch}{lccccccc}
Category & ggH cat 0 & ggH cat 1 & ggH cat 2 & ggH cat 3 & VBF cat 0 & VBF cat 1 & Combined \\
\hline
Observed limit (${10^{-4}}$) & ${<}0.53$ & ${<}0.85$  & ${<}1.45$ & ${<}10.54$  & ${<}2.08$ & ${<}3.96$  & ${<}0.44$   \\[1.0ex]
Expected limit (${10^{-4}}$) & ${<}0.82$ & ${<}0.82$  & ${<}1.32$ & ${<}7.12$  &  ${<}1.72$ & ${<}3.27$  & ${<}0.47$   \\[1.0ex]
\end{scotch}
\label{tab:limit_emu_125}
}
\end{table*}

The upper limit on \BHemSM is also interpreted as a constraint on the LFV Yukawa couplings ${Y_{\Pe\PGm}}$~\cite{Harnik:2012pb}. The LFV decay arises at tree level from the BSM Yukawa coupling, ${Y_{\Pe\PGm}}$. The decay width ${\Gamma(\HemSM)}$ can be written in terms of the Yukawa coupling as,
\begin{linenomath}
  \begin{equation}
    \Gamma(\HemSM) = \frac{\mh}{8\pi}(\abs{Y_{\Pe\PGm}}^2 + \abs{Y_{\PGm\Pe}}^2).
  \end{equation}
\end{linenomath}
The branching fraction \BHemSM assuming \HemSM is the only BSM contribution is given by,
\begin{linenomath}
  \begin{equation}
    \mathcal{B}(\HemSM) = \frac{\Gamma(\HemSM)}{\Gamma(\HemSM) + \Gamma_{\mathrm{SM}}}.
  \label{eq:yukawa_BR}
  \end{equation}
\end{linenomath}

\begin{figure}[!htb]
  \centering
  \includegraphics[width=0.48\textwidth]{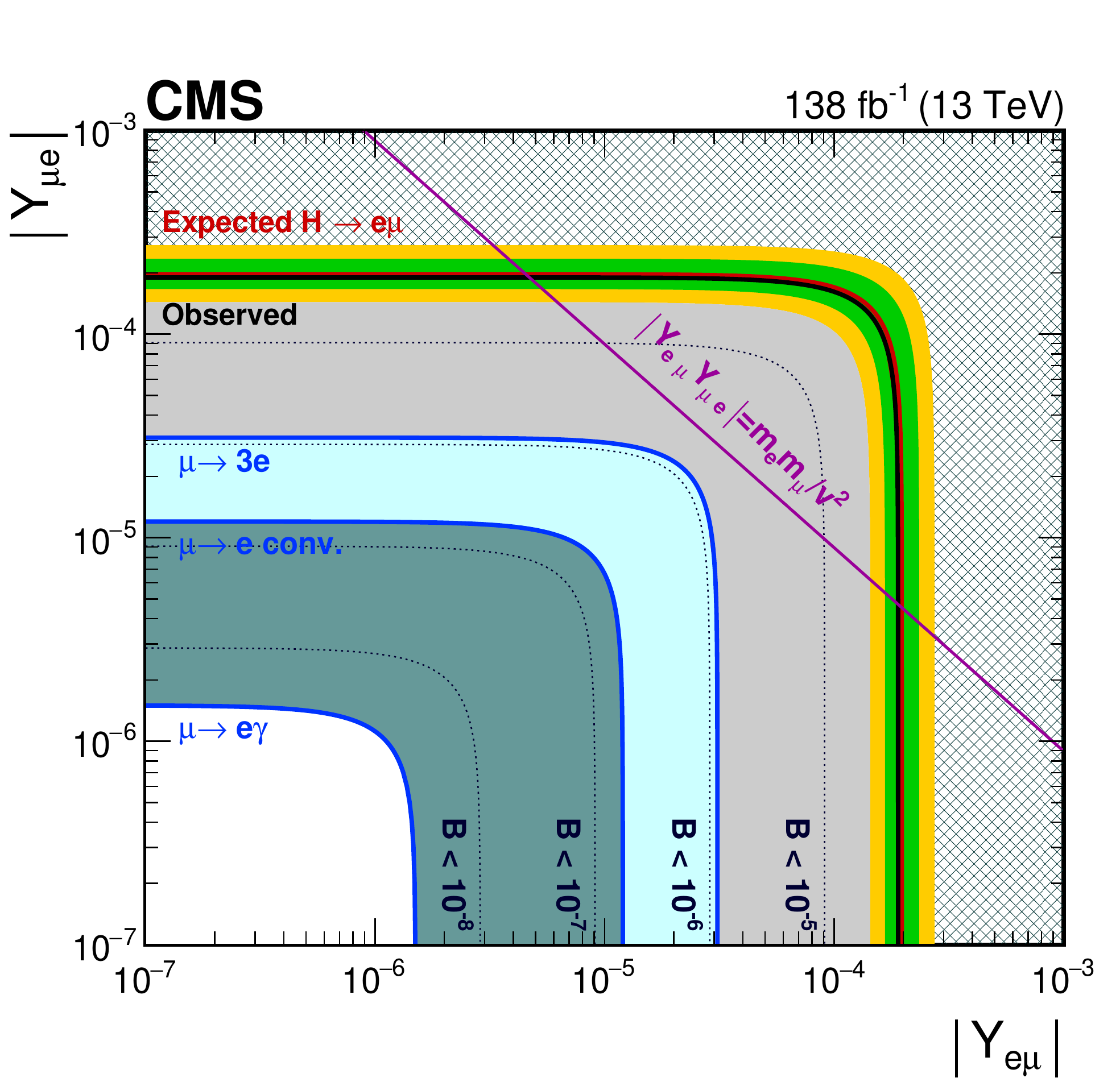}
	  \caption{Constraints on the lepton-flavor violating Yukawa couplings, ${\abs{Y_{\Pe\PGm}}}$ and ${\abs{Y_{\PGm\Pe}}}$. The observed (expected) limit in black (red) line is derived from the limit on \BHemSM in this analysis. The green (yellow) band indicates the one (two) standard deviation uncertainty in the expected limit. The hashed region is excluded by this direct search. Other shaded regions represent indirect constraints derived from the null searches for ${\PGm\to3\Pe}$ (gray)~\cite{SINDRUM:1987nra}, ${\PGm\to\Pe}$ conversion (light blue)~\cite{Kitano:2002mt}, and ${\PGm\to\Pe\Pgg}$ (dark green)~\cite{MEG:2016leq}. The flavor-diagonal Yukawa couplings, ${\abs{Y_{\Pe\Pe}}}$ and ${\abs{Y_{\PGm\PGm}}}$, are assumed to be at their SM values in the calculation of these indirect limits. The purple line is the theoretical naturalness limit of ${\abs{Y_{\Pe\PGm}Y_{\PGm\Pe}}\leq{m_{\Pe}}m_{\PGm}/v^2}$, where ${v}$ is the vacuum expectation value of the Higgs field. Dotted lines represent the corresponding constraints on ${\abs{Y_{\Pe\PGm}}}$ and ${\abs{Y_{\PGm\Pe}}}$ at upper limits on ${\BHemSM}$ at $10^{-5}, 10^{-6}, 10^{-7},$ and $10^{-8}$, respectively.}
  \label{fig:yukawa_limits}
\end{figure}

The decay width of \PH is assumed to be ${\Gamma_{\mathrm{SM}}=4.1\MeV}$ at ${\mh\approx125\GeV}$~\cite{Denner:2011mq}. The observed (expected) upper limit on the Yukawa coupling is evaluated to be ${\sqrt{\smash[b]{\Yemu^{2}+\Ymue^{2}}}<1.9\ (2.0)\times10^{-4}}$ at 95\% \CL. The result is illustrated in Fig.~\ref{fig:yukawa_limits}. 

\subsection{Results for additional Higgs bosons}
\begin{figure}[!htb]
  \centering
  \includegraphics[width=0.48\textwidth]{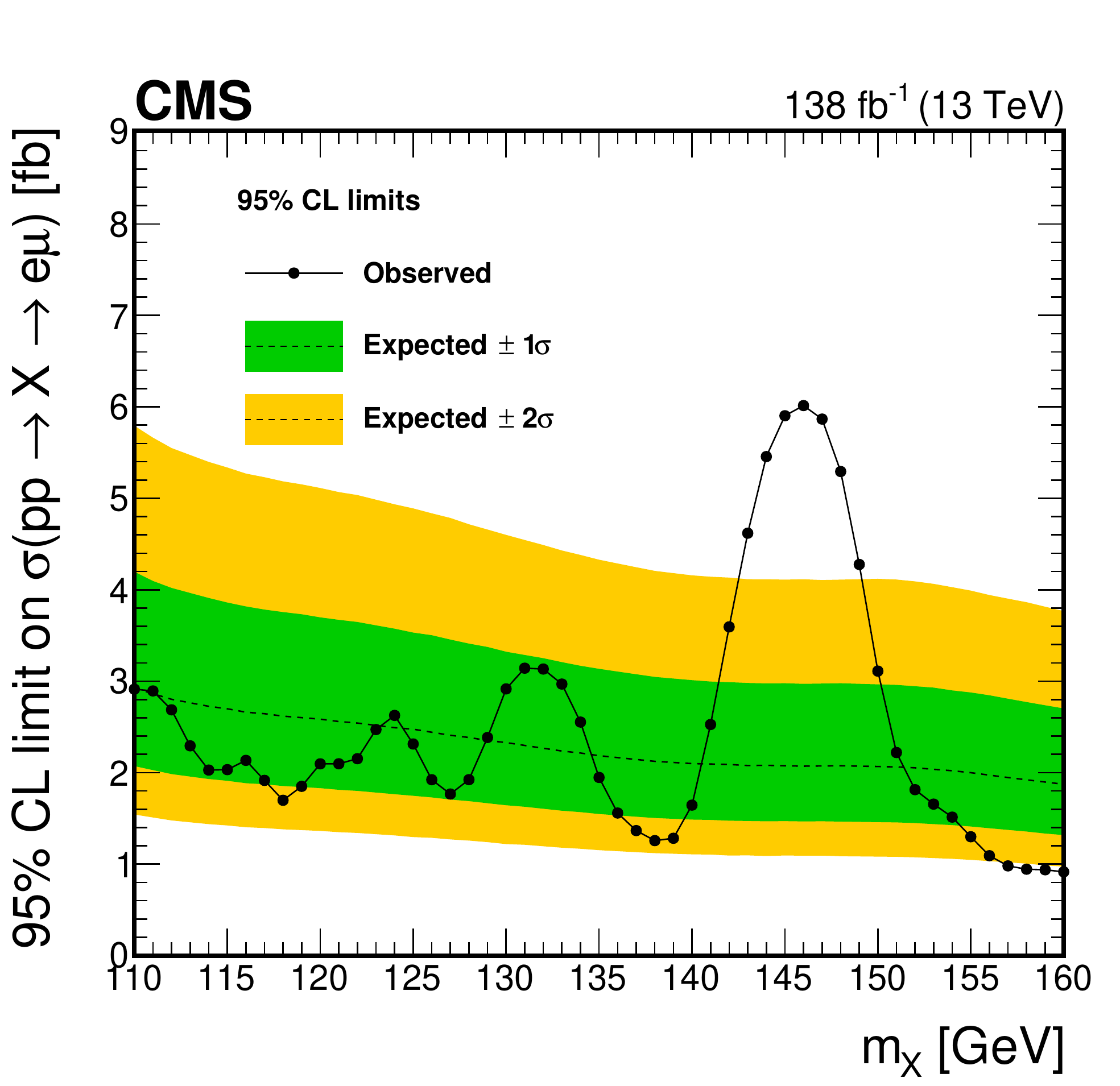}
  \includegraphics[width=0.48\textwidth]{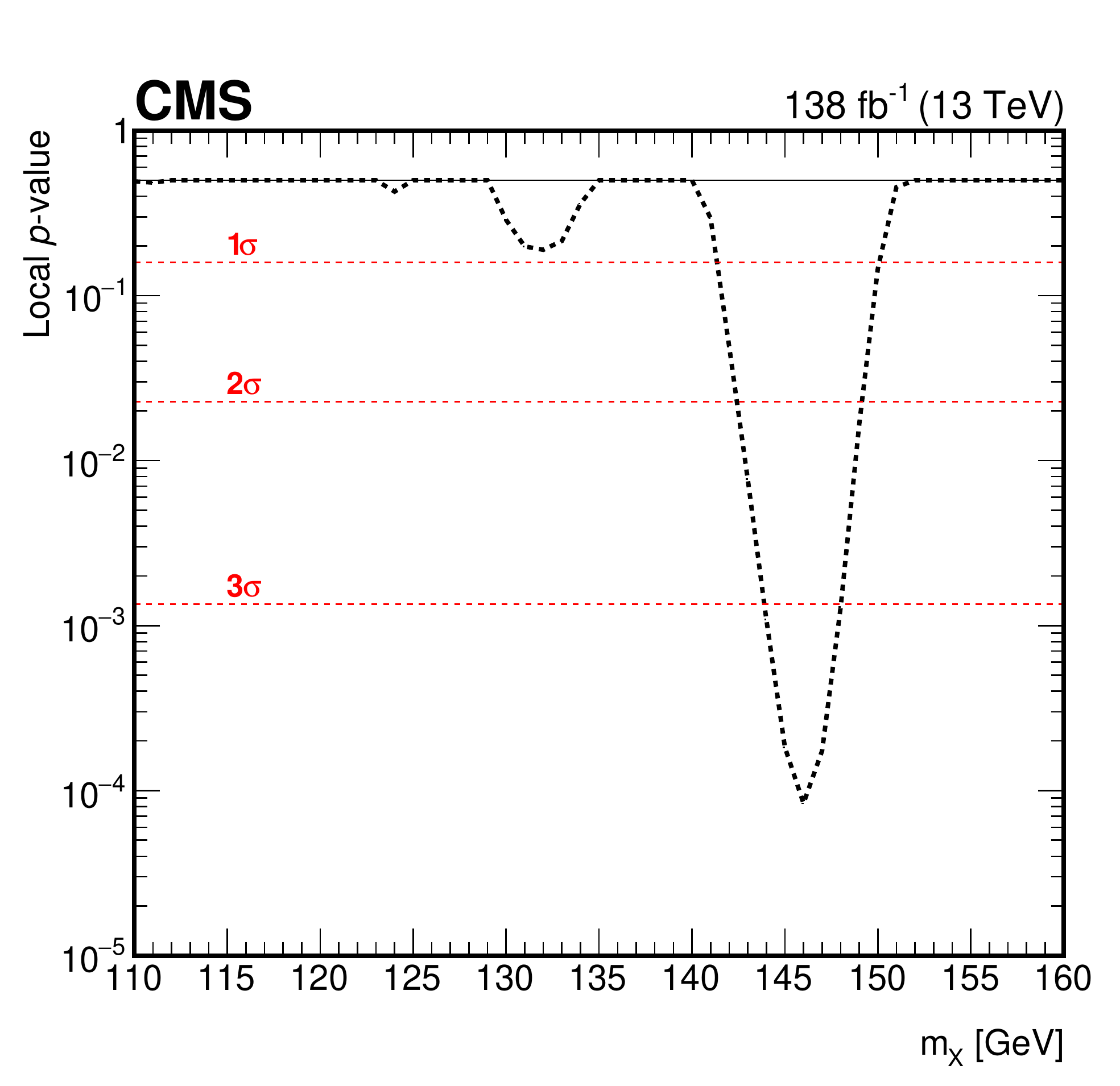}
  \caption{\cmsLLeft: the observed (expected) 95\% \CL upper limits on \sppXem as a function of the hypothesized \mx assuming the relative SM-like production cross sections of the \ggH and VBF production modes. \cmsRRight: the observed local $p$-values against the background-only hypothesis are shown as a function of the hypothesized \mx.}
  \label{fig:mass_scan}
\end{figure}

\begin{figure}[!htb]
  \centering
  \includegraphics[width=0.48\textwidth]{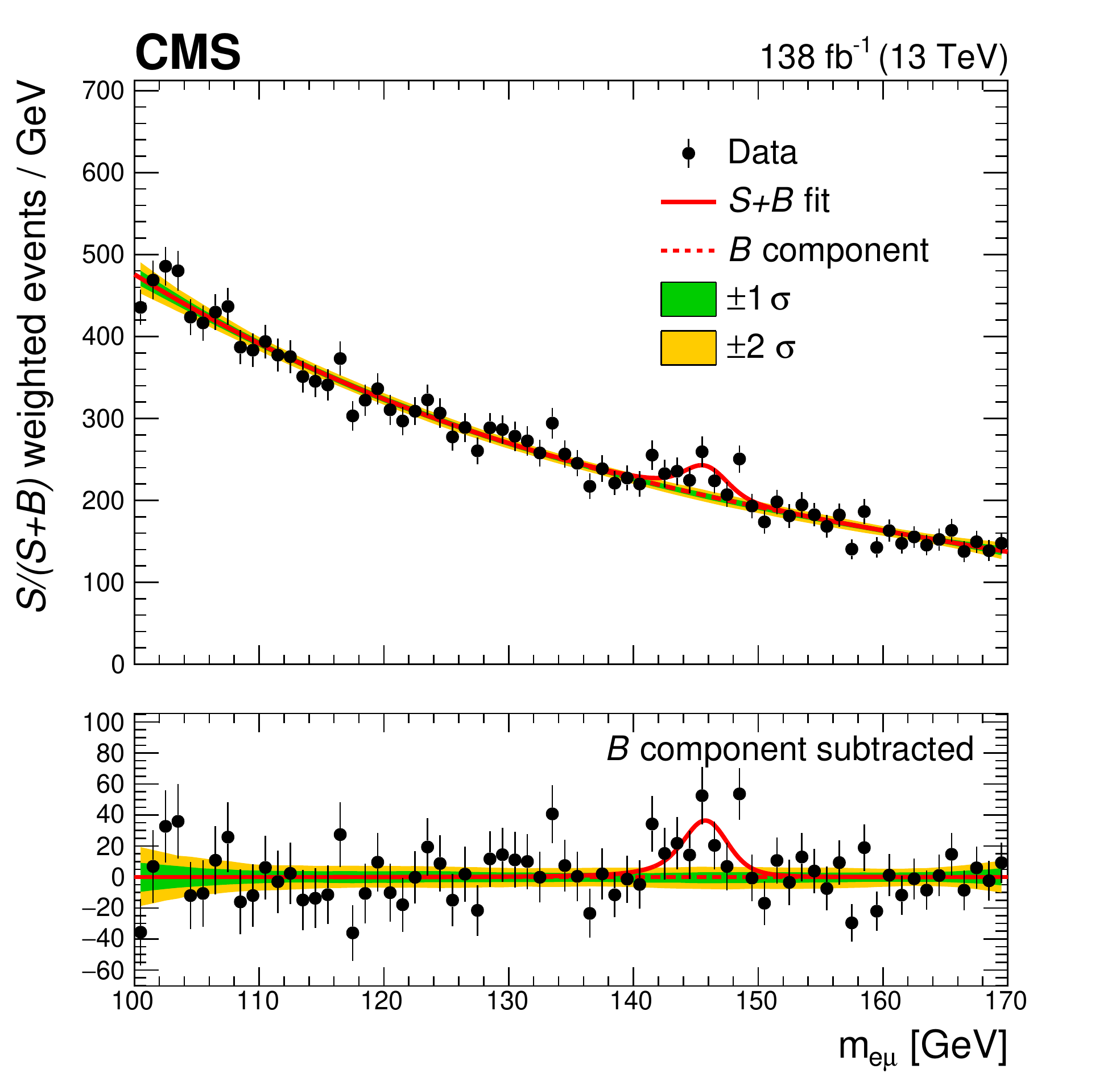}
  \caption{The \mem distribution of the observed data is shown, with the ${S+B}$ fit at ${\mx=146\GeV}$ in red solid line, and the background component of the fit in red dotted line. Events and fit in each category are weighted by ${S/(S+B)}$. The one and two standard deviation uncertainty bands of the background component are shown in green and yellow. The lower panel shows the residuals after subtracting the background component of the fit from data.}
  \label{fig:S+B_fits_comb_146}
\end{figure}

\begin{figure}[!htbp]
  \centering
  \includegraphics[width=0.48\textwidth]{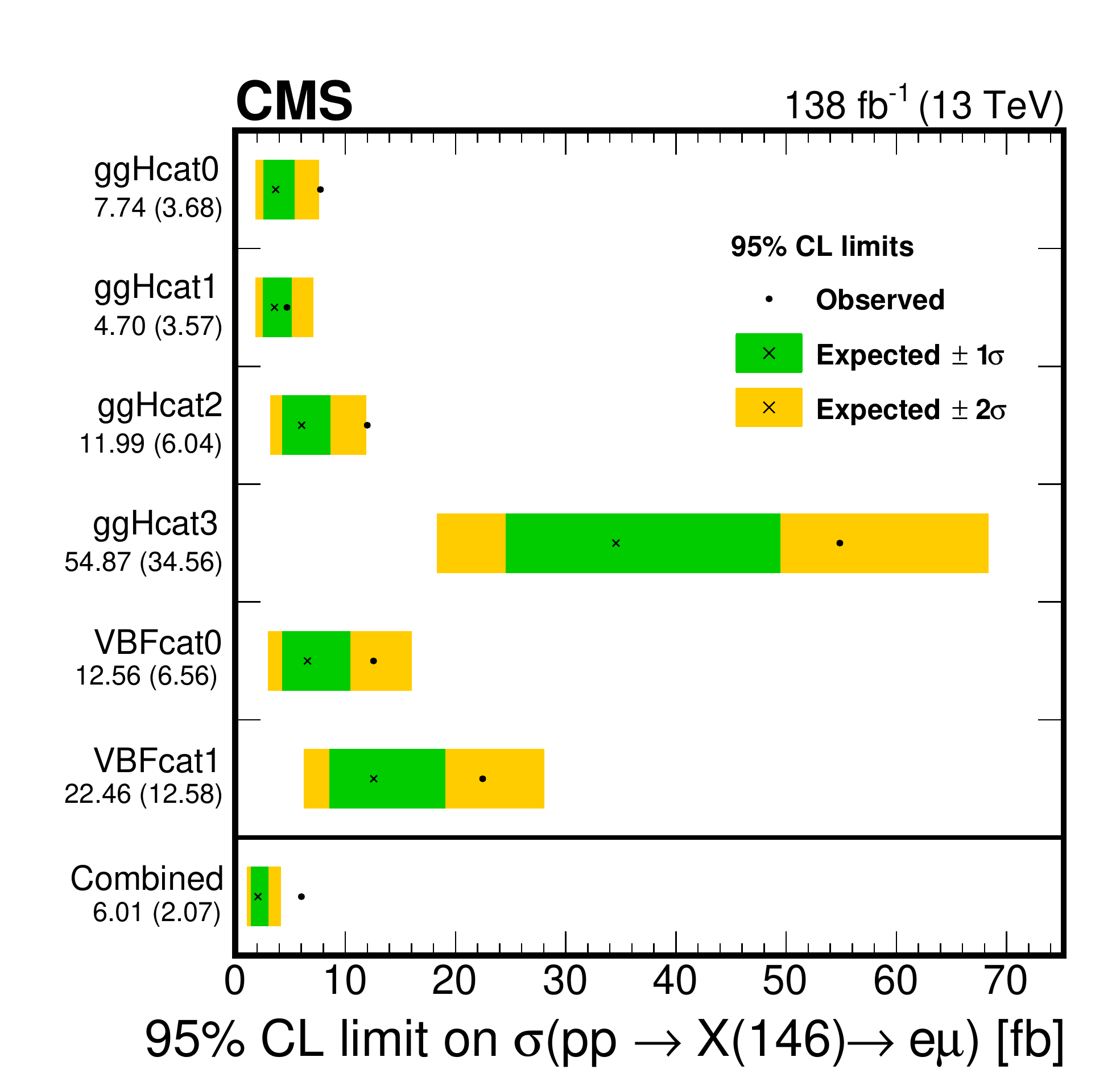}
  \caption{Observed (expected) 95\% \CL upper limits on \sppXemOFS for each individual analysis category (as shown in the left axis label) and for the combination of all analysis categories assuming the relative SM-like production cross sections of the \ggH and VBF production modes.}
  \label{fig:bf_emu}
\end{figure}

The observed (expected) upper limit at 95\% \CL on ${\sppXem}$ is plotted as a function of the hypothesized \mx in the range {110--160\GeV} on the \cmsLeft in Fig.~\ref{fig:mass_scan}, assuming the relative SM-like production cross sections of the \ggH and VBF production modes as evaluated in Ref.~\cite{deFlorian:2016spz}. An excess of events over the background-only hypothesis is observed at ${\mx\approx146\GeV}$. The corresponding ${S+B}$ fit combining all categories is shown in Fig.~\ref{fig:S+B_fits_comb_146}, where events in each category are weighted by ${S/(S+B)}$, where ${S}$ and ${B}$ are the fit number of signal and background events in that category. The observed and expected upper limits on \sppXemOFS at 95\% \CL per-category and combined are listed in Table~\ref{tab:bf_emu} and illustrated graphically in Fig.~\ref{fig:bf_emu}. The best fit of \sppXemOFS combining all analysis categories is ${3.89_{-1.08}^{+1.11}\text{(stat.)}_{-0.34}^{+0.57}\text{(syst.)}\mathrm{fb}}$, with the uncertainties dominated by the statistical uncertainties of the data. The best fit of \sppXemOFS per-category and combined with the corresponding local significance are also summarized in Table~\ref{tab:bf_emu}. Tabulated results are provided in the HEPData record~\cite{hepdata}. Such an excess, however, was not reported in a search of similar sensitivity for \HemSM carried out by the ATLAS experiment which covered the \mem range of the excess~\cite{Aad:2690381}.

\begin{table*}[htb!]
\centering
\caption{Observed (expected) 95\% \CL upper limits, best fit, and local significance in unit of standard deviation (${\sigma}$) of \sppXemOFS for each individual analysis category and for the combination of all analysis categories.}
\cmsTable{
\begin{scotch}{lccccccc}
Category & ggH cat 0 & ggH cat 1 & ggH cat 2 & ggH cat 3 & VBF cat 0 & VBF cat 1 & Combined \\
\hline
Observed limit (fb) & ${<}7.74$ & ${<}4.70$   & ${<}11.99$   & ${<}54.87$   & ${<}12.56$   &  ${<}22.46$ & ${<}6.01$ \\[1.0ex] 
Expected limit (fb) & ${<}3.68$ & ${<}3.57$   & ${<}5.04$   &  ${<}34.56$   & ${<}6.56$   &   ${<}12.58$ & ${<}2.07$ \\[1.0ex] 
Best fit (fb) & ${4.16_{-1.87}^{+2.10}}$       & ${1.30_{-1.78}^{+1.87}}$   & ${6.56_{-3.07}^{+3.25}}$   & ${23.46_{-17.31}^{+18.17}}$   & ${5.35_{-2.96}^{+3.92}}$   &  ${9.00_{-6.30}^{+7.46}}$ & ${3.89_{-1.13}^{+1.25}}$ \\[1.0ex] 
Local significance (${\sigma}$) & 2.3 & 0.7   & 2.2   & 1.4   & 2.1   &  1.5 & 3.8 \\[1.0ex] 
\end{scotch}
\label{tab:bf_emu}
}
\end{table*}

\section{Summary}
\label{sec:summary}
Searches for the lepton-flavor violation decay of \PH and \PX with a \mx in the range {110--160\GeV} have been performed in the \Pe{}\PGm final state in data collected by the CMS experiment. The data correspond to an integrated luminosity of {138\fbinv} of \Pp{}\Pp collisions at a center-of-mass energy of {13\TeV}. The observed (expected) upper limit on the branching fraction of the \PH decay \BHemSM is found to be ${4.4\ (4.7)\times10^{-5}}$ at 95\% confidence level, which is the most stringent direct limit set thus far. Upper limits on the cross sections of \ppXem are set in the \mx range {110--160\GeV} at 95\% confidence level. This is the first result of a direct search for \Xem, with \mx below twice the \PW boson mass. The largest excess of events over the expected background is observed with a local (global) significance of 3.8 (2.8) standard deviations at an invariant mass of the \Pe{}\PGm final state of around {146\GeV}. The observed significance of this excess is insufficient to draw any conclusions. More data will be needed to clarify the nature of the excess.

\begin{acknowledgments}
 We congratulate our colleagues in the CERN accelerator departments for the excellent performance of the LHC and thank the technical and administrative staffs at CERN and at other CMS institutes for their contributions to the success of the CMS effort. In addition, we gratefully acknowledge the computing centers and personnel of the Worldwide LHC Computing Grid and other centers for delivering so effectively the computing infrastructure essential to our analyses. Finally, we acknowledge the enduring support for the construction and operation of the LHC, the CMS detector, and the supporting computing infrastructure provided by the following funding agencies: BMBWF and FWF (Austria); FNRS and FWO (Belgium); CNPq, CAPES, FAPERJ, FAPERGS, and FAPESP (Brazil); MES and BNSF (Bulgaria); CERN; CAS, MoST, and NSFC (China); MINCIENCIAS (Colombia); MSES and CSF (Croatia); RIF (Cyprus); SENESCYT (Ecuador); MoER, ERC PUT and ERDF (Estonia); Academy of Finland, MEC, and HIP (Finland); CEA and CNRS/IN2P3 (France); BMBF, DFG, and HGF (Germany); GSRI (Greece); NKFIH (Hungary); DAE and DST (India); IPM (Iran); SFI (Ireland); INFN (Italy); MSIP and NRF (Republic of Korea); MES (Latvia); LAS (Lithuania); MOE and UM (Malaysia); BUAP, CINVESTAV, CONACYT, LNS, SEP, and UASLP-FAI (Mexico); MOS (Montenegro); MBIE (New Zealand); PAEC (Pakistan); MES and NSC (Poland); FCT (Portugal); MESTD (Serbia); MCIN/AEI and PCTI (Spain); MOSTR (Sri Lanka); Swiss Funding Agencies (Switzerland); MST (Taipei); MHESI and NSTDA (Thailand); TUBITAK and TENMAK (Turkey); NASU (Ukraine); STFC (United Kingdom); DOE and NSF (USA).
   
\hyphenation{Rachada-pisek} Individuals have received support from the Marie-Curie program and the European Research Council and Horizon 2020 Grant, contract Nos.\ 675440, 724704, 752730, 758316, 765710, 824093, 884104, and COST Action CA16108 (European Union); the Leventis Foundation; the Alfred P.\ Sloan Foundation; the Alexander von Humboldt Foundation; the Belgian Federal Science Policy Office; the Fonds pour la Formation \`a la Recherche dans l'Industrie et dans l'Agriculture (FRIA-Belgium); the Agentschap voor Innovatie door Wetenschap en Technologie (IWT-Belgium); the F.R.S.-FNRS and FWO (Belgium) under the ``Excellence of Science -- EOS" -- be.h project n.\ 30820817; the Beijing Municipal Science \& Technology Commission, No. Z191100007219010; the Ministry of Education, Youth and Sports (MEYS) of the Czech Republic; the Hellenic Foundation for Research and Innovation (HFRI), Project Number 2288 (Greece); the Deutsche Forschungsgemeinschaft (DFG), under Germany's Excellence Strategy -- EXC 2121 ``Quantum Universe" -- 390833306, and under project number 400140256 - GRK2497; the Hungarian Academy of Sciences, the New National Excellence Program - \'UNKP, the NKFIH research grants K 124845, K 124850, K 128713, K 128786, K 129058, K 131991, K 133046, K 138136, K 143460, K 143477, 2020-2.2.1-ED-2021-00181, and TKP2021-NKTA-64 (Hungary); the Council of Science and Industrial Research, India; the Latvian Council of Science; the Ministry of Education and Science, project no. 2022/WK/14, and the National Science Center, contracts Opus 2021/41/B/ST2/01369 and 2021/43/B/ST2/01552 (Poland); the Funda\c{c}\~ao para a Ci\^encia e a Tecnologia, grant CEECIND/01334/2018 (Portugal); the National Priorities Research Program by Qatar National Research Fund;  MCIN/AEI/10.13039/501100011033, ERDF ``a way of making Europe", and the Programa Estatal de Fomento de la Investigaci{\'o}n Cient{\'i}fica y T{\'e}cnica de Excelencia Mar\'{\i}a de Maeztu, grant MDM-2017-0765 and Programa Severo Ochoa del Principado de Asturias (Spain); the Chulalongkorn Academic into Its 2nd Century Project Advancement Project, and the National Science, Research and Innovation Fund via the Program Management Unit for Human Resources \& Institutional Development, Research and Innovation, grant B05F650021 (Thailand); the Kavli Foundation; the Nvidia Corporation; the SuperMicro Corporation; the Welch Foundation, contract C-1845; and the Weston Havens Foundation (USA).  
\end{acknowledgments}
\ifthenelse{\boolean{cms@external}}{\clearpage}{}

\bibliography{auto_generated}
\cleardoublepage \appendix\section{The CMS Collaboration \label{app:collab}}\begin{sloppypar}\hyphenpenalty=5000\widowpenalty=500\clubpenalty=5000
\cmsinstitute{Yerevan Physics Institute, Yerevan, Armenia}
{\tolerance=6000
A.~Hayrapetyan, A.~Tumasyan\cmsAuthorMark{1}\cmsorcid{0009-0000-0684-6742}
\par}
\cmsinstitute{Institut f\"{u}r Hochenergiephysik, Vienna, Austria}
{\tolerance=6000
W.~Adam\cmsorcid{0000-0001-9099-4341}, J.W.~Andrejkovic, T.~Bergauer\cmsorcid{0000-0002-5786-0293}, S.~Chatterjee\cmsorcid{0000-0003-2660-0349}, K.~Damanakis\cmsorcid{0000-0001-5389-2872}, M.~Dragicevic\cmsorcid{0000-0003-1967-6783}, A.~Escalante~Del~Valle\cmsorcid{0000-0002-9702-6359}, P.S.~Hussain\cmsorcid{0000-0002-4825-5278}, M.~Jeitler\cmsAuthorMark{2}\cmsorcid{0000-0002-5141-9560}, N.~Krammer\cmsorcid{0000-0002-0548-0985}, L.~Lechner\cmsorcid{0000-0002-3065-1141}, D.~Liko\cmsorcid{0000-0002-3380-473X}, I.~Mikulec\cmsorcid{0000-0003-0385-2746}, J.~Schieck\cmsAuthorMark{2}\cmsorcid{0000-0002-1058-8093}, R.~Sch\"{o}fbeck\cmsorcid{0000-0002-2332-8784}, D.~Schwarz\cmsorcid{0000-0002-3821-7331}, M.~Sonawane\cmsorcid{0000-0003-0510-7010}, S.~Templ\cmsorcid{0000-0003-3137-5692}, W.~Waltenberger\cmsorcid{0000-0002-6215-7228}, C.-E.~Wulz\cmsAuthorMark{2}\cmsorcid{0000-0001-9226-5812}
% ...
\par}
\vskip\cmsinstskip
\dag:~Deceased\\
$^{1}$Also at Yerevan State University, Yerevan, Armenia\\
$^{2}$Also at TU Wien, Vienna, Austria\\
% ...
\end{sloppypar}
\end{document}